\documentclass[paper,11pt]{JHEP}
\usepackage[centertags]{amsmath}
\usepackage{amsfonts} \usepackage{amssymb} \usepackage{amsthm}
\usepackage{graphicx}

\newcommand{\ii}{\mathrm{i}}
\newcommand{\ee}{\mathrm{e}}
\newcommand{\one}{{\rm 1\kern -.9mm l}}

\newcommand{\Tr}{\mathrm{Tr}\,}
\newcommand{\tr}{\mathrm{tr}\,}

\title{Classical solutions for exotic instantons?
}
\author{\parbox{11.5cm}{Marco Bill\`o$^1$, Marialuisa Frau$^1$, Laurent Gallot$^2$,
Alberto Lerda$^3$ and Igor Pesando$^1$}
\\
~\\
~\\
$^1$Dipartimento di Fisica Teorica, Universit\`a di Torino\\
and I.N.F.N. - sezione di Torino \\
Via P. Giuria 1, I-10125 Torino, Italy\\
\vspace{0.3cm}
$^2$
LAPTH\\
9, Chemin de Bellevue\\
74941 Annecy le Vieux Cedex, France \\
\vspace{0.3cm}
$^3$Dipartimento di Scienze e Tecnologie Avanzate, Universit\`a del Piemonte Orientale\\
and I.N.F.N. - Gruppo Collegato di Alessandria - sezione di Torino\\
Via V. Bellini 25/G, I-15100 Alessandria, Italy\\
\vspace{0.3cm}
\email{billo,frau,lerda,ipesando@to.infn.it; laurent.gallot@lapp.in2p3.fr }
}

\abstract{We consider the D7/D(--1) system in Type I$^\prime$ as a prototypical
``exotic'' brane instanton. With respect to systems such as the D3/D(--1) ones,
which correspond to gauge instantons in four dimensions, exotic systems lack the
bosonic mixed moduli $w$ of the ADHM construction, related to the instanton size
$\rho$, and their possible field-theoretical interpretation as classical
solutions is an important open question.
For the system at hand, we propose that it corresponds to the point-like limit
$\rho\to 0$ of the eight-dimensional so-called SO(8) instanton solution. This
configuration is a solution of the quartic term of the non-abelian D7 action,
{\it i.e.} the term which stays finite in the limit $\alpha'\to 0$ with $g_s$
fixed that preserves the D(--1) effects. As a necessary consistency condition,
we check that the next order term in the non-abelian effective action vanishes on
the SO(8) solution so that the limit we take is well-defined.
}
\preprint{DFTT/32/2008\\LAPTH 1297/08\\ESI-2100}

\begin{document}

\section{Introduction and motivations}
\label{sec:intro}
The construction of ``phenomenological'' models of particle physics based on D-branes embedded in supersymmetric string compactifications has become a major direction in the development
of String Theory (for reviews, see for instance \cite{Blumenhagen:2005mu,Blumenhagen:2006ci,Marchesano:2007de}).

In these D-brane scenarios non-perturbative corrections arise from D-instantons
and wrapped Euclidean branes. In principle all instantonic branes which can be
consistently included may contribute to the low energy effective action. If the
theory contains a gauge sector realized on D$(3+n)$-branes wrapped on a cycle
$\mathcal{C}$, then Euclidean branes E$(n-1)$ wrapped on $\mathcal{C}$
correspond to the instanton sectors of the gauge theory%
\footnote{The simplest case is represented by the D3/D(--1) system,
corresponding to $n=0$.} \cite{Witten:1995gx,Douglas:1995bn}. Other Euclidean
branes, for instance those wrapped on a cycle $\mathcal{C}^\prime\not=
\mathcal{C}$, do not possess this interpretation and have been referred to as
``exotic'' or ``stringy'' instantons; they have been investigated over the last
couple of years in a rapidly growing literature.
This interest was sparkled by the realization that exotic instantons might
provide couplings which are forbidden in perturbation theory but necessary
for phenomenological applications; for
instance, they have been pointed out as possible sources of neutrino masses
\cite{Blumenhagen:2006xt,Ibanez:2006da} or of certain Yukawa couplings in GUT
models \cite{Blumenhagen:2007zk,Kokorelis:2008ce}.
While the relation of ``ordinary'' instantonic branes to the field-theoretical description of instantons in supersymmetric gauge theories 
(as reviewed, for instance, in \cite{Dorey:2002ik}) has been clarified in detail \cite{Green:2000ke,Billo:2002hm,Billo:2006jm},
the possibility of a field-theoretic interpretation of the exotic instantonic
branes is still an open issue, and constitutes the main motivation of the
present work. 

In the ordinary cases, the spectrum and interactions of the moduli, {\it i.e.} of the physical excitations of open strings with at least one end-point on the instantonic branes, reproduce the ADHM construction of the moduli space of gauge theory instantons.
In particular, let us consider the NS sector of the open strings with one end on the D$(3+n)$ and the other end on the E$(n-1)$-brane. The world-sheet physicity condition has the form%
\footnote{In order to be explicit, we consider a toroidal orbifold situation, and assume that the branes in the internal space are distinguished by their relative angles or magnetizations.} 
\begin{equation}
 \label{vira}
L_0 - \frac 12 
= N_X + N_\psi +  \sum_{i=1}^3\frac{\theta_i}{2} = 0~,
\end{equation}
where $N_X$ and $N_\psi$ are the occupation numbers for the bosonic and
fermionic world-sheet oscillators, while the (positive) angles $2\pi\theta_i$
denote the twist eventually occurring in the three complex internal directions.
In writing (\ref{vira}) we have taken into account the $1/2$ contribution to the
zero-point energy from the four space-time directions, which are of
Neumann-Dirichlet type.
Ordinary E$(n-1)$-branes impose, in the internal space, the same boundary
conditions as the gauge D$(3+n)$-branes do. All twists $\theta^i$ therefore
vanish and the ground state $N_X=N_\psi=0$ is physical; being degenerate in the
non-compact directions, it corresponds to the moduli $w_{\dot\alpha}$ of the
ADHM construction. These bosonic mixed moduli enter in an essential way in the
ADHM constraints and, once these constraints are solved, they contain in
particular
the size $\rho$ of the instanton solution.

In the exotic cases, the E$(n-1)$-branes do not coincide with the gauge branes in the internal space, 
and the mixed strings have non-trivial twists $\theta^i$. Then the mass-shell condition (\ref{vira})
cannot be solved, and the bosonic mixed moduli $w_{\dot\alpha}$ are absent. This absence constitutes the main world-sheet hallmark of stringy instantons and has profound consequences. 
For ordinary D$(n+3)$/E$(n-1)$ systems the Higgs phase, in which a non-zero vacuum expectation value 
of the $w$'s is turned on and a bound state at threshold is formed, is the one which corresponds to the field theory instanton. This phase is not present in exotic configurations. 

Turning to the R sector of E$(n-1)$/E$(n-1)$ strings, there are fermionic anti-chiral moduli  $\lambda_{\dot\alpha}$ which,
in the moduli action, play the r\^ole of Lagrange multiplier for the fermionic ADHM constraints. In the exotic case the abelian part of such constraints, being proportional to the $w_{\dot\alpha}$ 
moduli, is absent and the abelian component of the $\lambda$'s represents a true fermionic zero-mode. To get non-zero correlators
it necessary to remove this zero-mode, for instance by appropriate orientifold projections
\cite{Argurio:2007qk,Argurio:2007vqa,Bianchi:2007wy},
or to lift it with closed string fluxes \cite{Blumenhagen:2007bn,Billo':2008sp,Billo':2008pg}
or by other mechanisms \cite{Petersson:2007sc,Ferretti:2009tz}.

In this paper we set out to identify the classical field configuration in a simple system of branes which from the world-sheet point of view shares the characteristics of exotic instantons, namely 
a system of D7 and D(--1)-branes in Type I$^\prime$ theory compactified on a torus $T_2$, as described in Section \ref{sec:system}.  In this system the D7/D(--1) strings have eight Neumann-Dirichlet directions; this corresponds to $\theta_1=\theta_2 = 1/2$ and
$\theta_3=0$ in the notation of (\ref{vira}). The bosonic moduli $w$ are therefore non-physical.
On the other hand, the orientifold projects out the dangerous fermionic zero-mode $\lambda_{\dot\alpha}$, and thus the D(--1)'s can contribute to the effective action.
The gauge theory living on the D7's is an eight-dimensional theory:
however, one could further compactify down to four dimensions,
to obtain a stringy instanton configuration composed of wrapped D7's and D(--1)'s. 

The eight-dimensional D7/D(--1) system is interesting by itself in many respects;
together with the T-dual situation in which D9/E1 systems on $T_2$ are considered,
it provides an important testing ground for the heterotic/Type I duality.
Indeed, the BPS-saturated $t_8 F^4$ terms in the effective Lagrangian, 
exactly known from the heterotic side, receive non-perturbative corrections from D(--1)'s 
on the type I side, that would be important to determine also from an explicit D-instanton calculus.
The heterotic result has already been reorganized so as to match the expected 
structure of D-instanton terms \cite{Bachas:1997mc}-\nocite{Kiritsis:1997hf,Bachas:1997xn,Foerger:1998kw,Bianchi:1998vq}\cite{Gava:1999ky},
but the explicit integration, at all instanton numbers, over the moduli space of the
D(--1)'s has not yet been performed. The preliminary step of describing the spectrum of moduli of the D7/D(--1) system and their action was already done in \cite{Gutperle:1999xu}, whose analysis substantially matches the one reported here in Sections \ref{subsec:spectrum} and \ref{subsec:modact}.
In the present paper our focus is on the identification of the classical solution corresponding to the D-instanton, rather than on the explicit moduli space integration; however the form of the field-dependent moduli action that we derive in Section \ref{subsec:scaling} represents a convenient starting point for the computation of the non-perturbative effective action. 

It is natural to expect that the D7/D(--1) system 
corresponds to some eight-dimensional classical configuration analogous to the gauge instanton
in four dimensions, whose properties and main features are spelled out in Sections \ref{sec:7-1} and \ref{sec:oneloop}.
In particular, considering both tree-level and one-loop amplitudes with part of
their boundary on the D(--1)'s, we argue that the classical configuration
represents an instanton of the quartic part of the Non-Abelian Born-Infeld (NABI)
action of the D7-branes. We then proceed in Section \ref{sec:eight} to review
the properties of known eight-dimensional instantons available in the
literature. The study of generalizations of gauge instantons has a
rather long history, and several types of configurations have been individuated
\cite{Corrigan:1982th}\nocite{Fubini:1985jm,Corrigan:1984si,Grossman:1984pi,Grossman:1989bb,
Duff:1990wu,Harvey:1990eg,Duff:1991sz,Gunaydin:1995ku}-\cite{Bak:2002aq},
depending on which feature of the four
dimensional case is extended to eight dimensions and taken as a definition of instantons.
We argue that the configuration which carries the appropriate symmetries to be associated 
to a D(--1) inside a stack of D7-branes is the supersymmetric version of the 
so-called SO(8) instanton of Grossman et al. \cite{Grossman:1984pi,Grossman:1989bb}
\begin{equation}
 \label{grossF}
F_{\mu\nu}= -\frac{2 \rho^2}{(x^2+\rho^2)^2}\,\gamma_{\mu\nu}~.
\end{equation}
Here the $\gamma_{\mu\nu}$ are the chiral gamma-matrices with two indices in $d=8$, while the field strength $F_{\mu\nu}$ is in the adjoint of SO(8), the smallest gauge group for which a configuration with non-trivial fourth Chern class can be obtained. 
Notice that in the Type I$^\prime$ system we consider, we have exactly $N=8$ D7 branes at each orientifold plane, thus supporting an SO(8) gauge theory%
\footnote{In the heterotic/Type I context, the SO(8) instanton was already 
considered in \cite{Minasian:2001ib} in the search for a configuration of Type I 
representing the heterotic string.}.

Reconsidering the moduli spectrum and the absence of emission diagrams for the gauge field in the D7/D(--1) system related to the absence of the bosonic $w$ moduli, we show in Section \ref{sec:vacuum} that actually it is the limit $\rho\to 0$ of the SO(8) instanton that corresponds to the D(--1). We show then that the NABI action evaluated on the SO(8) instanton in this 
limit reduces to its quartic term; this is not at all
trivial, as it requires that all higher order terms of the NABI action vanish identically already at finite $\rho$. In Section \ref{sec:vanishing} we will check this explicitly for the quintic term,
which is of order ${\alpha'}^3$ with respect to the quadratic one. In the literature, different expressions of the quintic term in the NABI action have been derived, starting from different guiding principles. We use the expression obtained in \cite{Collinucci:2002ac} from the requirement that the NABI action should admit an off-shell supersymmetric extension. Other expressions in the literature \cite{Refolli:2001df}\nocite{Koerber:2001uu,Grasso:2002wb}-\cite{Barreiro:2005hv} are equivalent to this one only up to terms proportional to the Yang-Mills equations of motion; however 
the SO(8) instanton is not a solution of the Yang-Mills equations in eight dimensions, but rather of those that follow from the quartic action. Therefore, the fact that these alternative forms of the quintic action do not vanish on the SO(8) instanton is not unexpected. We think that this observation can be relevant to the problem of determining which form of the NABI action should be used to weight off-shell configurations. 
It is not possible at the moment to do further checks, since no higher
correction has been determined according to the principle of
supersymmetrizability used in \cite{Collinucci:2002ac}; nevertheless the fact
that the quintic term vanishes on the configuration (\ref{grossF}) makes the
conjecture that all higher terms vanish as well very plausible.

\section{The D7/D(--1) system in Type I$^\prime$}
\label{sec:system}

We consider Type IIB string theory compactified on a 2-torus ${T}_2$
and modded out by 
\begin{equation}
\Omega = \omega \,(-1)^{F_L}\,{\mathcal I}_2
\label{orientifoldparity}
\end{equation}
where $\omega$ is the world-sheet parity, $F_L$ is the left-moving world-sheet fermion
number, and ${\mathcal I}_2$ is the inversion along the two directions of $T_2$.
The resulting theory is an unoriented string model with sixteen
supercharges, called Type I$^\prime$. 

The action of $\Omega$ has four fixed-points on the torus where four O7-planes are placed.
A local cancellation of the R-R tadpoles produced by these O7-planes requires to place 
at each fixed-point eight D7-branes. In the following we will concentrate on one
fixed-point and study the gauge theory produced by the massless modes of the
unoriented open strings attached to the eight D7-branes that are located there.
This is an eight-dimensional gauge theory with $\mathcal N=1$ supersymmetry and gauge group
$\mathrm{SO}(8)$.
A quick way to see this is to observe that the vertex operators for the massless bosonic
modes on $N$ D7 branes contain (in the 0 superghost picture) the following terms%
\footnote{Here, for simplicity, we only write terms that do not depend on the
world-sheet fermions. The symbols $\partial_\tau$ and $\partial_\sigma$ denote, 
respectively, the world-sheet tangent and normal derivatives;
$k$ is the open string momentum, and Greek (Latin) indices label 
the longitudinal (transverse) directions of the D7 branes with Euclidean signature,
{\it i.e.} $\mu,\nu,...=1,...,8$ and $m,n,...=9,10$.} :
$V_A \sim A_\mu\,\partial_\tau X^\mu\,\ee^{\ii k_\nu X^\nu}$ 
and $V_\phi \sim \phi_m\,\partial_\sigma X^m\,\ee^{\ii k_\nu X^\nu}$. They
both change sign under $\Omega$. If we assume that the action $\gamma(\Omega)$
of the orientifold projection on the Chan-Paton factors is
just a transposition, namely
\begin{equation}
 A_\mu \to \gamma(\Omega)\,A_\mu\gamma(\Omega)^{-1}= {}^{\mathrm{t}}A_\mu
\quad , \quad
\phi_m \to \gamma(\Omega)\,\phi_m \gamma(\Omega)^{-1}= {}^{\mathrm{t}}\phi_m~,
\label{gammaomega7}
\end{equation}
we easily conclude that the vertex operators $V_A$ and $V_\phi$ survive the
orientifold projection only if $A_\mu$ and $\phi_m$ are $N\times N$ anti-symmetric matrices.
In our case ($N=8$), we therefore have an eight-dimensional vector $A_\mu$ and
two real scalars $\phi_m$ which transform in the adjoint representation of $\mathrm{SO}(8)$.
A similar analysis in the fermionic massless sector of these unoriented strings leads to an eight-dimensional chiral
fermion $\Lambda^{\alpha}$ plus
its anti-chiral conjugate $\Lambda_{\dot\alpha}$, which also
transform in the adjoint representation of $\mathrm{SO}(8)$. Altogether, the fields
\begin{equation}
 \big\{A_\mu, \Lambda^\alpha, \phi_m\big\}
\label{vecmul}
\end{equation}
form the $\mathcal N=1$ vector multiplet in $d=8$ in the adjoint representation of $\mathrm{SO}(8)$
and can be assembled into a superfield as follows
\begin{equation}
 \Phi(x,\theta) = \phi(x) + \sqrt{2}\,\theta\Lambda(x) +\frac{1}{2}\,
\theta\gamma^{\mu\nu}\theta\,F_{\mu\nu}(x) + \ldots
\label{Phi}
\end{equation}
where $\phi=(\phi_9+\ii\phi_{10})/\sqrt{2}$ (see Appendix \ref{appa} for our conventions
on $\gamma$ matrices, etc.).

The part of the effective action which depends only on the gauge field strength $F_{\mu\nu}$
and its covariant derivatives is the NABI action for D7 branes, which
can be organized in a series of contributions with increasing powers of $\alpha'$
and contains the following terms \cite{Tseytlin:1986ti,Tseytlin:1997csa}:
\begin{eqnarray}
S_{\mathrm{D7}} & = &\frac{1}{8\pi g_s}\int \!d^8x \,\mathrm{Tr}
\left[\frac{1}{(2\pi\sqrt{\alpha'})^4}\,F_{\mu\nu}F^{\mu\nu} 
- \frac{1}{3\,(2\pi)^2}\, t_8\,F^4 \right]
+ \frac{\alpha'}{g_s} \int \!d^8x\, \mathcal{L}_{(5)}(F,DF)+\cdots\nonumber\\
{\phantom{\frac{1}{8\pi g_s}}}& = & S_{\mathrm{YM}} + S_{(4)} + S_{(5)} + \cdots,
\label{action0}
\end{eqnarray}
where $g_s$ is the string coupling constant and
\begin{eqnarray}
& &\Tr \big(t_8 F^4\big) \equiv  \frac{1}{16}\,t_8^{\mu_1\mu_2\cdots\mu_7\mu_8} \,\Tr \big(F_{\mu_1\mu_2}\cdots F_{\mu_7\mu_8}\big)
\label{f4}\\
& = &\Tr\Big(F_{\mu\nu}F^{\nu\rho}F^{\lambda\mu}F_{\rho\lambda} 
+\frac{1}{2}\,F_{\mu\nu}F^{\rho\nu}F_{\rho\lambda}F^{\mu\lambda}
-\frac{1}{4}\,F_{\mu\nu}F^{\mu\nu}F_{\rho\lambda}F^{\rho\lambda}
-\frac{1}{8}\,F_{\mu\nu}F_{\rho\lambda}F^{\mu\nu}F^{\rho\lambda}\Big)~.
\nonumber
\end{eqnarray}
In the Yang-Mills action $S_{\mathrm{YM}} 
= 1/(2 g_{\mathrm{YM}}^2)\int \!d^8x\, \mathrm{Tr}(F^2)$, there appears a dimensionful
coupling
\begin{equation}
 \label{gym8}
{g_{\mathrm{YM}}^2} \equiv {4\pi g_s (2\pi\sqrt{\alpha'})^4}~,
\end{equation}
while the quartic action, which we write as
\begin{equation}
 \label{S40}
S_{(4)} = -\frac{1}{4!\lambda^4} \int d^8x\, \mathrm{Tr}\big(t_8 F^4\big)~, 
\end{equation}
bears in front a dimensionless coupling constant
\begin{equation}
 \label{lambdadef}
{\lambda^4} \equiv {4\pi^3 g_s}~.
\end{equation}
Finally, the last term in (\ref{action0}) is an $O(\alpha')$ contribution whose expression and 
r\^ole will be discussed in Section \ref{sec:vanishing}.

As mentioned in the Introduction, we are interested in studying non-perturbative effects
that are induced by $k$ D(--1)-branes located at the same fixed point of the D7 branes.
The D-instantons are sources for the scalar field $C_0$ of the closed string
R-R sector, and their
classical action is 
\begin{equation}
 S_{\mathrm{D(-1)}} = \frac{2\pi}{g_s}-2\pi\ii\,C_0 = -2\pi\ii\,\tau~,
\label{smod1}
\end{equation} 
where 
\begin{equation}
 \tau=C_0+\frac{\ii}{g_s}
\label{axiondil}
\end{equation}
is the standard axion-dilaton combination.
The resulting D7/D(--1)-brane system is 1/2-BPS and stable since the number of directions with mixed Dirichlet-Neumann boundary conditions is eight. Even if there will be important
differences, it is useful to analyze this D7/D(--1)-brane system in analogy with the 
more familiar D3/D(--1)-brane system.

In the ``standard'' case of D3/D(--1) systems, the sector with $k$ D(--1) branes corresponds to the sector with instanton number $k$ of the four dimensional gauge theory living on the D3-branes
\cite{Witten:1995gx,Douglas:1995bn}. 
Indeed the part of the D3 action quadratic in the gauge field reads%
\footnote{We assume for simplicity that the R-R field $C_0$ is constant along the D3 world-volume.}
\begin{equation}
 \label{D3f2}
S_{(2)} = \frac{1}{8\pi g_s} \int d^4x \,\mathrm{Tr}\big(F^2\big) -2\pi\ii\, C_0 \, c_{(2)}~,
\end{equation}
where $c_{(2)}$ is the second Chern number
\begin{equation}
 \label{chern2}
c_{(2)}= \frac{1}{8\pi^2}\int {\mathrm Tr} \big(F\wedge F\big)~.
\end{equation}
On an instanton configuration which saturates the Bogomolny bound with instanton number $k=c_{(2)}$, namely when
\begin{equation}
 \label{bogym}
\frac 12 \int d^4x \,\mathrm{Tr}\big(F^2\big) = 8\pi^2\, c_{(2)}~,
\end{equation}
the action $S_{(2)}$ reduces to that of $c_{(2)}$ D-instantons:
\begin{equation}
 \label{D3inst}
S_{(2)} = -2\pi\ii\,c_{(2)}\,\tau = c_{(2)}\,S_{\mathrm{D(-1)}}~.
\end{equation}

Instanton configurations satisfying (\ref{bogym}) can be described through the ADHM construction. 
They preserve the space-time $\mathrm{SO}(4)$ invariance,
and in a supersymmetric context represent 1/2-BPS states;
both these features are to be expected if they must correspond to a D3/D(--1) configuration.
In fact, the moduli of the D3/D(--1)-brane system and their action account precisely for the ADHM construction of the instanton moduli space \cite{Green:2000ke,Billo:2002hm}.
The interpretation of the D(--1)'s as a classical configuration of action (\ref{D3f2}) 
is further confirmed by the relation between instantonic one-loop amplitudes 
and gauge threshold corrections to the quadratic action, as originally found in \cite{Abel:2006yk,Akerblom:2006hx}. Indeed, as discussed in 
\cite{Billo:2007sw,Billo:2007py}, both instantonic one-loop amplitudes 
and gauge threshold corrections determine the one-loop running of the gauge coupling constant
that is obtained by expanding around two different backgrounds, respectively an instanton and a constant field. 
Finally, it was shown in \cite{Billo:2002hm} that the D(--1)'s act as sources for
the classical profile of the solution through the emission of the gauge field
from a ``mixed'' disk  
with its boundary attached partly to the D(--1)'s and partly to the D3's. 

Let us now consider a system with D7 and D(--1) branes. The part of the Born-Infeld action for D7 branes that is \emph{quartic} in the gauge fields is given in (\ref{S40}) and, after
adding to it the quartic term from the the Wess-Zumino part, 
it can be rewritten in the following form
\begin{equation}
 \label{S4nuova}
S_{(4)} = -\frac{1}{4!\,4\pi^3 g_s} \int d^8x\, \mathrm{Tr}\big(t_8 F^4\big) 
-{2\pi\ii\,C_0}\,c_{(4)}
\end{equation}
where $c_{(4)}$ is the fourth Chern number
\begin{equation}
 c_{(4)} = \frac{1}{4!\,(2\pi)^4}\,\int {\mathrm Tr} \big(F\wedge
F\wedge F\wedge F\big)
= \frac{1}{4!\,(2\pi)^4}\,\int d^8x\, \mathrm{Tr}\big(\epsilon_8 F^4\big)~,
\end{equation}
with
\begin{equation}
\Tr \big(\epsilon_8 F^4\big) \equiv
\frac{1}{16}\,\epsilon_8^{\mu_1\mu_2\cdots\mu_7\mu_8} \,\Tr \big(F_{\mu_1\mu_2}\cdots F_{\mu_7\mu_8}\big)~.
\end{equation}
If the gauge field $F$ satisfies
\begin{equation}
 \int d^8x\, \mathrm{Tr}\big(t_8 F^4\big) = -\frac{1}{2}
\int d^8x\, \mathrm{Tr}\big(\epsilon_8 F^4\big)  = -\frac {4!}{2}(2\pi)^4\, c_{(4)}
\label{t8Fepsilon8F}
\end{equation}
then the quartic action (\ref{S4nuova}) becomes
\begin{equation}
\label{S4SD-1}
 S_{(4)} = -2\pi\ii\,c_{(4)}\,\tau = c_{(4)}\, S_{\mathrm{D(-1)}}~,
\end{equation}
that is the action of $c_{(4)}$ D-instantons.

This argument suggests that, in analogy to the D3/D(--1) case, $k$ D(--1)-branes inside a stack of D7-branes should correspond to some classical ``instanton'' configuration that satisfies (\ref{t8Fepsilon8F}) with $k=c_{(4)}$. 
In the rest of the paper, we will pursue this line of inquiry, 
analyzing in turn the various aspects of such a relation that we recapitulated above in the case
of D3/D(--1) systems. Along the way we will point out analogies and differences.

\section{Spectrum and interactions in the D7/D(--1) system}
\label{sec:7-1}
We begin by considering the tree-level aspects of the D7/D(--1) system; in particular
we describe the moduli spectrum corresponding to the allowed vertex operators 
and the action resulting from their disk interactions.

\subsection{Moduli spectrum}
\label{subsec:spectrum}
The spectrum in the D7/D(--1) system consists of neutral moduli, associated to
open strings with both end-points on the D-instantons ($(-1)/(-1)$ strings),
and of charged moduli, associated to open strings stretching between the gauge 
and the instantonic branes ($7/(-1)$ strings). 

\paragraph{(--1)/(--1) strings} This is the neutral sector since it comprises states that
do not transform under the gauge group. The physical zero-modes are easily obtained
by dimensionally reducing the $\mathcal N=1$ supersymmetric gauge theory from ten to
zero dimensions. Using an ADHM-inspired notation, we denote
the bosonic fields as $a_\mu$ and $\chi_m$, where the distinction between the two is made by
the presence of the D7-branes. Both $a_\mu$ and $\chi_m$ are $k\times k$ matrices and have 
canonical dimensions of (length)$^{-1}$.
In order to implement the orientifold projection, we observe that since all directions have Dirichlet-Dirichlet boundary conditions,
the vertex operator for $a_\mu$ (in the 0 superghost picture) is proportional to
$\partial_\sigma X^\mu$, and not to $\partial_\tau X^\mu$, so that it is even under $\Omega$. On the other hand, the vertex operator for $\chi_m$ is proportional to $\partial_\sigma X^m$ and is odd
because of the reflection $\mathcal I_2$ of the coordinates $X^m$ transverse to the
O7-planes.
Furthermore, the consistency with the orientifold action 
(\ref{gammaomega7}) on the D7-branes requires \cite{Gimon:1996rq} that on the
Chan-Paton factors of the (--1)/(--1) strings $\Omega$ acts also as a transposition, {\it i.e.}
\begin{equation}
 a_\mu \to \gamma(\Omega)\,a_\mu\gamma(\Omega)^{-1}= {}^{\mathrm{t}}a_\mu
\quad , \quad
\chi_m \to \gamma(\Omega)\,\chi_m \gamma(\Omega)^{-1}= {}^{\mathrm{t}}\chi_m~.
\label{gammaomega-1}
\end{equation}
Combining the behavior of the operator part of the vertex operators with that of their Chan-Paton
factors, we easily conclude that $a_\mu$ is a $k\times k$ matrix transforming in the symmetric representation of $\mathrm{SO}(k)$, while $\chi_m$ is a $k\times k$ matrix in the anti-symmetric
(adjoint) representation of $\mathrm{SO}(k)$.

Let us now turn to the fermionic sector. Adopting again an ADHM-inspired notation, we denote
the chiral and anti-chiral fermionic moduli as $M^\alpha$ and $\lambda_{\dot\alpha}$. 
They both have dimensions of (length)$^{-\frac32}$ and correspond to the
following vertex operators (in the $-\frac12$ superghost picture)%
\footnote{In these expressions we understand factors of $(2\pi\alpha')^{\frac34}$
which are needed to make $V_M$ and $V_\lambda$ dimensionless.} 
\begin{equation}
 V_M = M^\alpha\,S_{\alpha} S_- \,\ee^{-\frac12\varphi}
\quad,\quad
V_\lambda = \lambda_{\dot\alpha}\, S^{\dot\alpha} S^+ \,\ee^{-\frac12\varphi}~.
\label{vertMlambda}
\end{equation}
Here $\varphi$ is the bosonic field appearing in the bosonization of the superghost
system, $S_\alpha$ and $S^+$ are the chiral spin fields in the first eight and last two directions respectively, while $S^{\dot\alpha}$ and $S_-$ are their anti-chiral counterparts.
The two combinations $S_{\alpha} S_-$ and $S^{\dot\alpha} S^+$ appearing in (\ref{vertMlambda})
arise upon splitting the (anti-chiral) spinor representation of $\mathrm{SO}(10)$ 
under $\mathrm{SO}(8) \times \mathrm{SO}(2)$ as required by the presence of the D7 branes.
On these spin fields we can represent the orientifold projection $\Omega$ 
as minus the chirality operator in the transverse directions to the D7-branes, or equivalently
as the chirality operator along the eight longitudinal directions%
\footnote{The choice of signs in these chirality operators corresponds to choose between instantons and anti-instantons; here and in the following we choose signs that, in our conventions,
are appropriate for instanton-like configurations.}.
Thus $S_{\alpha} S_-$ is even while $S^{\dot\alpha} S^+$ is odd under $\Omega$.
On the other hand, in analogy with (\ref{gammaomega-1}), the Chan-Paton factors transform as
\begin{equation}
 M^\alpha \to \gamma(\Omega)\,M^\alpha\gamma(\Omega)^{-1}= {}^{\mathrm{t}}M^\alpha
\quad , \quad
\lambda_{\dot\alpha} \to \gamma(\Omega)\,\lambda_{\dot\alpha} \gamma(\Omega)^{-1}= {}^{\mathrm{t}}\lambda_{\dot\alpha}~,
\label{gammaomega-1f}
\end{equation}
so that $V_M$ and $V_\lambda$ survive the orientifold projection only if $M^\alpha$ is
a symmetric $k\times k$ matrix of $\mathrm{SO}(k)$ and $\lambda_{\dot\alpha}$ is 
an antisymmetric $k\times k$ matrix of $\mathrm{SO}(k)$.

In the one D-instanton case ($k=1$), sometimes also called the $\mathrm O(1)$ instanton, 
the moduli $\chi_m$ and $\lambda_{\dot\alpha}$ are projected out and the only neutral zero-modes
that survive are $a_\mu$ and $M^\alpha$. These represent the bosonic and fermionic Goldstone modes 
of the (super)translations of the D7-branes world-volume that are broken by the D-instanton and
thus can be identified with the bosonic and fermionic coordinates $x_\mu$ and $\theta^\alpha$
of the eight-dimensional superspace. More precisely, we have
\begin{equation}
 x_\mu= (2\pi\alpha')\, a_\mu\quad,\quad
\theta^\alpha = (2\pi\alpha')\, M^\alpha
\label{xtheta}
\end{equation}
where the factors of $\alpha'$ have been introduced to give $x_\mu$ and $\theta^\alpha$
the appropriate dimensions.

\paragraph{7/(--1) strings} This is the charged sector that accounts for open
strings stretching between the D7-branes and the D-instantons. Since there are
eight directions with mixed Dirichlet-Neumann boundary conditions,
in the NS sector it is not possible to construct vertex operators of conformal weight
one and thus, according to (\ref{vira}), 
there are no physical bosonic moduli in the spectrum. This is to be
contrasted with what
happens in the D3/D(--1) system where, instead, one finds bosonic moduli $w_{\dot\alpha}$ and 
$\bar w_{\dot\alpha}$ that are related to the gauge instanton size. 
On the other hand, the absence of charged bosonic moduli is the distinctive
feature of the ``exotic'' instanton configurations mentioned in the Introduction, and thus, 
at least from this point of view, our D7/D(--1) system is in the same class.

The fermionic R sector instead is not empty. In fact,
we can construct the following vertex operators for the fermionic moduli
$\mu$ and $\bar\mu$ (of canonical dimensions of (length)$^{-\frac32}$)
\begin{equation}
 V_\mu = \mu\,\Delta\,S^+\,\ee^{-\frac12\varphi}\quad,\quad
V_{\bar\mu} = \bar\mu\,\bar\Delta\,S^+\,\ee^{-\frac12\varphi}
\label{Vmu}
\end{equation}
where $\Delta$ and $\bar \Delta$ are the (bosonic) twist operators for the eight directions with
mixed boundary conditions%
\footnote{In (\ref{Vmu}) we have understood factors of
$(2\pi\alpha')^{\frac34}$.}. They are conformal fields with conformal dimension
$1/2$ and refer to the two orientations of the string stretching between
the D7-branes and the D-instantons.
Fixing the chirality for the spin field in the last two directions 
is a GSO projection; the choice of the positive chirality made in (\ref{Vmu})
is, in our conventions, the appropriate one 
for instanton-like configurations%
\footnote{We note in particular that the vertex
operators (\ref{Vmu}) are mutually local with the supercurrents of the supersymmetry conserved
by the D7/D(--1) system.}. Under the orientifold action the two vertex operators
$V_\mu$ and $V_{\bar\mu}$ are mapped into each other; indeed
\begin{equation}
 V_\mu\,\xrightarrow{\Omega} -\, {}^{\mathrm{t}}\mu\,\bar\Delta\,S^+\,\ee^{-\frac12\varphi}
\quad,\quad
V_{\bar\mu}\,\xrightarrow{\Omega} -\, {}^{\mathrm{t}}{\bar\mu}\,\Delta\,S^+\,\ee^{-\frac12\varphi}
\label{vomega}
\end{equation}
where the minus sign is due to the fact that $\Omega$ acts as
minus the chirality operator on the spin fields in the last two directions.
Thus $\mu$ and $\bar\mu$ are not independent of each other but are related as
follows
\begin{equation}
 \bar\mu = - \,{}^{\mathrm{t}}\mu
\label{mubarmut}
\end{equation}
Remember that $\mu$ and $\bar\mu$ are, respectively, $N\times k$ and $k\times N$ matrices
(with $N=8$ in our specific case); the above identification is therefore consistent
with this structure. 

We conclude by stressing again that the content of the charged sector of the D7/D(--1) moduli
spectrum is similar to that of the charged sector 
in exotic instanton configurations (see for example  
\cite{Blumenhagen:2006xt,Ibanez:2006da,Argurio:2007vqa,Bianchi:2007wy}),
or to that of the flavored sector in D3/D(--1) systems 
that realize $\mathcal N=1$ or $\mathcal N=2$ SQCD models (see for example
\cite{Billo:2007sw,Billo:2007py} for details). Therefore, we can proceed like in these
cases and work with rescaled moduli \cite{Billo:2002hm} 
\begin{equation}
 \mu' = \frac{2\pi\alpha'}{\sqrt{g_s}}\,\mu
\label{mu'}
\end{equation}
which carry dimensions of (length)$^{\frac12}$.

\subsection{Moduli action}
\label{subsec:modact}

The D-instanton moduli action can be derived by computing scattering amplitudes
of moduli fields on disks with at least part of their boundary on
the D(--1)-branes; we refer to \cite{Green:2000ke,Billo:2002hm} for details and
we recall here only the most significant results for our purposes.

First of all, the contribution $S_0$ to the moduli action of a disk with no moduli insertion is
\cite{Polchinski:1994fq}
\begin{equation}
S_0 = \frac{2\pi k}{g_s}= \frac{k}{2\pi^2\alpha'^2g_0^2}
\label{S0}
\end{equation}
where $g_0$ is the Yang-Mills coupling constant in zero dimensions and $k$ is the multiplicity of
the disk boundary.
Note that $S_0/k$ is nothing else than the (topological) normalization of any disk amplitude 
with D(--1) boundary conditions (denoted as $C_0$ in \cite{Billo:2002hm}): clearly this normalization is always the same since the D(--1)-branes are objects which exist independently 
of whether the surrounding space-filling branes that support the gauge theory are D7 or D3-branes.

Next, we have the contribution $S_1$ of the neutral moduli which is obtained from
disk amplitudes involving the vertex operators of the (--1)/(--1) strings. 
The result of these computations is formally similar to the one discussed 
in \cite{Billo:2002hm}, and for our present case it is explicitly given by
\begin{equation}
\begin{aligned}
S_1 = &\,\frac{1}{g_0^2}\,\mathrm{tr}\Big\{ \ii\,\lambda_{\dot\alpha}
\gamma_\mu^{\dot\alpha\beta} [a^\mu,M_\beta] -
\frac{\ii}{\sqrt{2}}\,\lambda_{\dot\alpha}[\chi,\lambda^{\dot\alpha}] 
- \frac{\ii}{\sqrt{2}}\,M^\alpha[\bar\chi,M_\alpha] 
\\ &
-\frac{1}{4}[a_\mu,a_\nu][a^\mu,a^\nu] - [a_\mu,\chi][a^\mu,\bar\chi]
+\frac{1}{2}[\chi,\bar\chi]^2\Big\}
\end{aligned}
\label{S1}
\end{equation}
where $\chi\equiv(\chi_9+\ii\chi_{10})/\sqrt{2}$, $\bar\chi\equiv(\chi_9-\ii\chi_{10})/\sqrt{2}$.

Finally, we have the contribution of the charged moduli that
arises from mixed disc amplitudes involving the boundary changing vertex operators
(\ref{Vmu}). In our case this contribution is simply
\begin{equation}
 S_2 = -\frac{\ii}{g_0^2}\,\mathrm{tr}\big( {}^{\mathrm{t}}\mu \,\mu \,\chi \big)
= -\ii \pi\,\mathrm{tr}\big({}^{\mathrm{t}}\mu'\,\mu'\,\chi\big)
\label{S2}
\end{equation}
where in the last step we have introduced the rescaled moduli defined in (\ref{mu'}).

In the one-instanton case $(k=1)$ there are drastic simplifications; indeed both  
terms (\ref{S1}) and (\ref{S2}) vanish, and after
including the contribution of the tadpole for the R-R scalar $C_0$, one
is simply left with the D-instanton action $S_{\mathrm{D(-1)}}$ given in (\ref{smod1}).
If one considers also the interactions among the charged moduli and the 7/7 strings, this
action gets replaced by
\begin{equation}
S_{\mathrm{D(-1)}}(\Phi)= -2\pi\ii\,\tau+\ii\pi\,{}^{\mathrm{t}}\mu'\,\Phi(x,\theta)\,\mu'
\label{smod12}
\end{equation}
where $\Phi$ is the superfield defined in (\ref{Phi}). The second term in (\ref{smod12}) can be
understood by computing a string amplitude on mixed disks with two charged moduli and a scalar
field of the 7/7 sector, plus its supersymmetric partners.

The moduli action of our D7/D(--1) system is much simpler than the
corresponding one for the D3/D(--1) systems, even in the one-instanton case:
the number of moduli is smaller, there are no bosonic charged
moduli like $w_{\dot\alpha}$ or
$\bar w_{\dot\alpha}$ and no ADHM-like constraints. This last
feature makes a crucial difference: indeed, the D7 and D(--1)-branes do not
form a bound state, while by imposing the ADHM constraints in the
D3/D(--1) case it is possible to obtain a bound state at threshold.

\subsection{Integration over moduli and scaling to the quartic theory}
\label{subsec:scaling}

The previous analysis shows that in the case of a single  D-instanton the moduli space is parametrized
by the following zero-modes: 
\begin{equation}
\mathcal{M}_{k=1} =\big\{x_\mu, \theta^\alpha, \mu'\big\}
\label{mod1}
\end{equation}
and that the superspace coordinates appear in the moduli action only through the dynamical gauge fields
on the D7-branes.
{From} the scaling dimensions of the moduli (\ref{mod1}) we easily prove that the
dimension of the measure for the moduli space integral corresponding to
this configuration is given by
\begin{equation}
 \Big[d\mathcal{M}_{k=1}\Big] = \ell_s^{(n_x-\frac12n_\theta-\frac12n_{\mu'})}=
\ell_s^{\frac{8-N}{2}}
\label{dimmeas}
\end{equation}
where we have used the string length $\ell_s=\sqrt{2\pi\alpha'}$ as unit of measure, and denoted by
$n_x$, $n_\theta$ and $n_{\mu'}$ the numbers of $x$, $\theta$ and $\mu'$'s. This calculation 
can be generalized to the case of an arbitrary number $k$ of D-instantons and the result is%
\footnote{Notice that for arbitrary $k$ it is necessary to introduce suitable auxiliary moduli that disentangle the
quartic interactions in the action (\ref{S1}) of the neutral moduli.
It turns out that the only dimensionful
new parameters that are introduced in this way are seven auxiliary moduli
$D_i$ ($i=1,\ldots,7$) which have dimensions of (length)$^{-2}$ and transform in the anti-symmetric
representation of $\mathrm{SO}(k)$. These $D_i$'s play in the present case the same
role that the three auxiliary moduli $D_c$ play in the D3/D(--1) system \cite{Billo:2002hm}.}
\begin{equation}
 \Big[d\mathcal{M}_{k}\Big] = \ell_s^{\frac{k(8-N)}{2}}~.
\label{dimmeas1}
\end{equation}
Therefore, in order to compensate for these dimensions the overall normalization of the
instanton measure must contain a factor of $\ell_s^{\frac{k(N-8)}{2}}$ which, together with the
classical action (\ref{S0}), implies that the non-perturbative contributions of this 
D7/D(--1) system are proportional to
\begin{equation}
 \ell_s^{\frac{k(N-8)}{2}}\,\ee^{-\frac{2\pi k}{g_s}}~.
\label{prefactor}
\end{equation}

In general, the contribution to the D7-brane effective action from the sector with $k$ D-instantons 
is obtained, as usual, by integrating over the moduli with an exponential weight given by the field-dependent moduli action:
\begin{equation}
\label{npea}
S_{\mathrm{eff}}^{(k)}(\Phi)
=\,\ell_s^{\frac{k(N-8)}{2}}\int d^8x\, d^8\theta \int d{\widehat{\mathcal{M}}}_{k}~
\ee^{2\pi\ii\tau k+S_1+S_2 +\ii\pi \tr ({}^{\mathrm{t}}\mu^\prime \Phi \mu^\prime)}~,
\end{equation}
where $\widehat{\mathcal{M}}_{k}$ denotes the centred moduli ({\it i.e.} all moduli but $x$ and $\theta$) and the last term in the exponent is the generalization of the field-dependent part of (\ref{smod12}) to arbitrary $k$.
For $N=8$, by simple dimensional analysis the non-perturbative effective action (\ref{npea})
takes the schematic form
\begin{equation}
 \label{npea1}
S_{\mathrm{eff}}^{(k)}(\Phi) = c_k \,\ee^{2\pi\ii\tau k}\int d^8x\, d^8\theta \,[\Phi^4]_{(k)} 
\end{equation}
where $c_k$ are numerical coefficients and the symbol $[\Phi^4]_{(k)}$ denotes a gauge-invariant combination quartic in the superfield $\Phi$ arising in the $k$-instanton sector.
The explicit expression of such invariant and the coefficients $c_k$ are determined by the integration over the centred moduli $\widehat{\mathcal{M}}_{k}$. Performing this integration and translating the result in heterotic
variables would give a stringent test of the duality between the Type I$^\prime$ theory and the heterotic string compactified on $T_2$ \cite{Bachas:1997mc}-\nocite{Kiritsis:1997hf,Bachas:1997xn,Foerger:1998kw,Bianchi:1998vq}\cite{Gava:1999ky}. We leave this task for future work.

As is clear from (\ref{dimmeas1}), for $N=8$ the measure $d\mathcal{M}_k$ is always dimensionless. 
The $N=8$ case resembles thus the $\mathcal N=4$ SYM theory in $d=4$,
where the ADHM instanton measure is dimensionless for any instanton number. 
Since this property of the ADHM measure is
strictly related to the vanishing of the $\beta$-function and hence to the conformal 
invariance at the quantum level of the $\mathcal N=4$ SYM theory, we are naturally
led to conjecture that the dimensionless character of $d\mathcal{M}_k$ 
points to the fact that the D-instantons inside the D7-branes of Type I$^\prime$
describe non-perturbative configurations of an eight-dimensional conformal
gauge theory. The relevant action for this theory cannot be the usual SYM
action, which in $d=8$ is not even conformal at the classical level, 
but an obvious candidate exists:
it is the quartic action $S_{(4)}$ given in (\ref{S4nuova})
plus its supersymmetric completion. This term appears naturally in the expansion
in powers of $\alpha'$ of the non-Abelian Born-Infeld (NABI) action of D7-branes, as shown
in (\ref{action0}); it has a coupling constant $\lambda$, given in (\ref{lambdadef}),
which is dimensionless, and it is conformal at the classical level.
On this basis we can therefore argue that the correct scaling with $\alpha'$ to be considered for
a field theory interpretation of the D7/D(--1)-brane system is the one in which
the dimensionless coupling $\lambda$ of the $t_8 F^4$ term, 
and not the Yang-Mills coupling $g_{\mathrm{YM}}$, is held fixed.

\section{One-loop amplitudes}
\label{sec:oneloop}
In order to support the interpretation of the D(--1)-branes as a
particular configuration of the gauge theory living on the D7-branes,
and in particular of the quartic action (\ref{S4nuova}), we perform
two calculations%
\footnote{For simplicity in this section we set $C_0=0$.}.
First, in
Section \ref{subsec:oneloopinst} we compute the one-loop vacuum energy
due to the open strings with at least one-end point on $k$
D-instantons and find agreement with the calculation of the scaling
dimension of the measure on the moduli space of the D7/D(--1) system
given in (\ref{dimmeas1}); 
second, in Section \ref{subsec:bckg} we compute the one-loop vacuum
energy of $N$ D7-branes in a constant background field $\mathcal F$
and, after extracting the contribution proportional to $\mathcal F^4$
we find a perfect matching with the instantonic one-loop amplitude
previously determined.  
{From} these calculations we are able to extract also the
renormalization properties of the quartic coupling $\lambda$ defined
in (\ref{lambdadef}) and show that it does not run for $N=8$.

\subsection{Type I$^\prime$ instantonic amplitudes}
\label{subsec:oneloopinst}
The one-loop vacuum energy $\Gamma_{\mathrm{D(-1)}}$ of $k$ D-instantons in presence of $N$ D7-branes
is given by
\begin{equation}
\Gamma_{\mathrm{D(-1)}}= Z_{\mathrm{D(-1)}} + Z_{\mathrm{D(-1)/D7}} + Z_{\mathrm{D7/D(-1)}}  
\label{oneloop}
\end{equation}
where $Z_{\mathrm{D(-1)}}$ is the contribution of the unoriented open strings attached to
the $k$ D(--1)-branes, while $Z_{\mathrm{D(-1)/D7}}$ and $Z_{\mathrm{D7/D(-1)}}$ are the contributions
of the open strings stretching between the D-instantons and the D7-branes or viceversa.

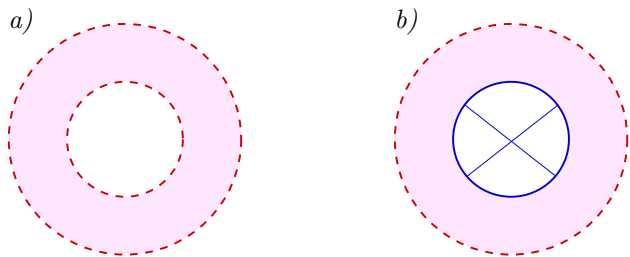
\begin{figure}[htb]
 \begin{center}
 \begin{picture}(0,0)%
\includegraphics{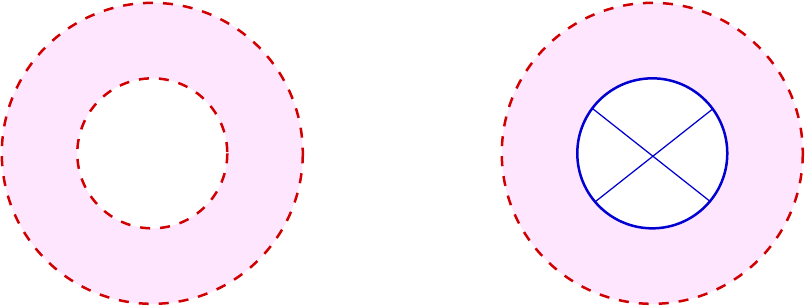}%
\end{picture}%
\setlength{\unitlength}{1579sp}%
\begingroup\makeatletter\ifx\SetFigFontNFSS\undefined%
\gdef\SetFigFontNFSS#1#2#3#4#5{%
  \reset@font\fontsize{#1}{#2pt}%
  \fontfamily{#3}\fontseries{#4}\fontshape{#5}%
  \selectfont}%
\fi\endgroup%
\begin{picture}(9809,3871)(211,-3179)
\put(226,389){\makebox(0,0)[lb]{\smash{{\SetFigFontNFSS{10}{12.0}{\familydefault}{\mddefault}{\updefault}\emph{a)}}}}}
\put(6226,389){\makebox(0,0)[lb]{\smash{{\SetFigFontNFSS{10}{12.0}{\familydefault}{\mddefault}{\updefault}\emph{b)}}}}}
\end{picture}%
 \end{center}
 \caption{\emph{a)} The annulus with D(--1) boundary conditions, 
indicated by the dashed line.  \emph{b)} The M\"obius diagram obtained with the 
 insertion of the O7 projection $\Omega$.}
 \label{fig:annulus_moebius}
\end{figure}
The one-loop vacuum energy $Z_{\mathrm{D(-1)}}$ has the following schematic form
\begin{equation}
Z_{\mathrm{D(-1)}} = \int_0^\infty \frac{dt}{2t} \,\,
\mathrm{Tr}_{\mathrm{D(-1)}}\left(\frac{1 + \Omega}{2}\,P_{\mathrm{GSO}}\, q^{L_0}\right)
=\frac{1}{2}\Big(\mathcal A_{\mathrm{D(-1)}}+\mathcal M_{\mathrm{D(-1)}}\Big)
~,
 \label{z1loop}
\end{equation}
where $P_{\mathrm{GSO}}$ is the GSO projector, $q=\exp(-2\pi t)$ and the trace 
$\mathrm{Tr}_{\mathrm{D(-1)}}$ is computed over all
open string states with D(--1) boundary conditions. As usual, we can decompose
the unoriented amplitude $Z_{\mathrm{D(-1)}}$ as the sum of an annulus
diagram, depicted in Fig. \ref{fig:annulus_moebius}$a)$,
\begin{equation}
\mathcal A_{\mathrm{D(-1)}} = \int_0^\infty \frac{dt}{2t} \,\,
\mathrm{Tr}_{\mathrm{D(-1)}}\left(P_{\mathrm{GSO}}\, q^{L_0}\right)~,
 \label{annulus}
\end{equation}
and of a M\"obius diagram, represented in Fig. \ref{fig:annulus_moebius}$b)$,
\begin{equation}
\mathcal M_{\mathrm{D(-1)}} = \int_0^\infty \frac{dt}{2t} \,\,
\mathrm{Tr}_{\mathrm{D(-1)}}\left(\Omega\,P_{\mathrm{GSO}} \,q^{L_0}\right)~,
 \label{moebius}
\end{equation}
which we now evaluate in turn. 

The calculation of the annulus amplitude is completely standard and the result is
\begin{equation}
  \mathcal A_{\mathrm{D(-1)}} = \frac{k^2}{2}
\!\int\frac{dt}{2t}
\left[\Big(\!\!-\!2\pi\frac{\theta_3(0|\ii t)}{\theta_1'(0|\ii t)}\Big)^4
- \Big(\!\!-\!2\pi\frac{\theta_4(0|\ii t)}{\theta_1'(0|\ii t)}\Big)^4
-\Big(\!\!-\!2\pi\frac{\theta_2(0|\ii t)}{\theta_1'(0|\ii t)}\Big)^4
\right]\mathcal W(t)~.
\label{A11I}
\end{equation}
Here $\theta_a(z|\ii t)$ ($a=1,\ldots,4$) are the Jacobi $\theta$-functions, 
$\theta_1'(0|\ii t)\equiv\partial_z\theta_1(z|\ii t)\big|_{z=0}$, and 
$\mathcal W(t)$ represents the sum over the winding modes in the two compact 
transverse directions, given by
\begin{equation}
 \mathcal W(t) = \sum_{(r_1,r_2)\in \mathbb Z^2} \ee^{-2\pi t\,\frac{|r_1+r_2U|^2 T_2}{U_2}}
\label{W}
\end{equation}
with $U=U_1+\ii U_2$ and $T=T_1+\ii T_2$ being, respectively, the complex and K\"ahler structure
of the 2-torus. The three terms in the square brackets of (\ref{A11I}) are the contributions of
the non-zero modes in the NS, NS$(-1)^F$ and R sectors; furthermore
the average over the two orientations of the $(-1)/(-1)$ strings has been explicitly 
taken into account in writing the overall coefficient. 
Due to the \emph{aequatio identica satis abstrusa} obeyed by the
$\theta$-functions, we have
\begin{equation}
\mathcal A_{\mathrm{D(-1)}} = 0~.
\end{equation}

Also the contribution of the M\"obius diagram (\ref{moebius}) can be easily computed, but some care 
must be used in evaluating the contribution of the R$(-1)^F$ sector. Indeed,
since the orientifold reflection is due to orientifold O7 planes, the fermionic zero-modes of the
$(-1)/(-1)$ strings contribute to the odd spin structure \cite{Billo:1998vr}; 
in particular, they give opposite results
for the two string orientations, as explained in detail in \cite{Billo:2007sw,Billo:2007py}.
Taking all this into account, we have
\begin{equation}
\begin{aligned}
  \mathcal M_{\mathrm{D(-1)}} = -\frac{k}{2}
\int\frac{dt}{2t}\,&\left\{\frac{1}{2}
\left[16\,\Big(\frac{\theta_4(0|\ii t+\frac12)}{\theta_2(0|\ii t+\frac12)}\Big)^4
- 16\,\Big(\frac{\theta_3(0|\ii t+\frac12)}{\theta_2(0|\ii t+\frac12)}\Big)^4
-1
\right]\mathcal W(t)\right.\\
&\left.+\frac{1}{2}\left[16\,\Big(\frac{\theta_4(0|\ii t+\frac12)}{\theta_2(0|\ii t+\frac12)}
\Big)^4- 16\,\Big(\frac{\theta_3(0|\ii t+\frac12)}{\theta_2(0|\ii t+\frac12)}\Big)^4
+1
\right]\mathcal W(t)\right\}~.
\end{aligned}
\label{M1I}
\end{equation}
The two lines above correspond to the two different orientations, the three terms in the square
brackets refer, respectively, to the NS, NS$(-1)^F$ and R$(-1)^F$ sectors, and the factors of 16
are introduced to compensate the $2^4$ factors brought by $(\theta_2)^4$.
Summing the various terms in (\ref{M1I}) and using the abstruse
identity, we have
\footnote{When interpreted in the closed string channel,
the M\"obius amplitude (\ref{M1I}) receives contributions from the exchange
between the D(--1) boundary states and the crosscaps located at all the four orientifold fixed points.
Thus, from this point of view this expression does not possess a truly local interpretation.}
\begin{equation}
 \mathcal M_{\mathrm{D(-1)}} = 8k\int\frac{d t}{2t} \,\mathcal W(t)~.
\label{M1I1}
\end{equation}

\begin{figure}[htb]
 \begin{center}
 \begin{picture}(0,0)%
\includegraphics{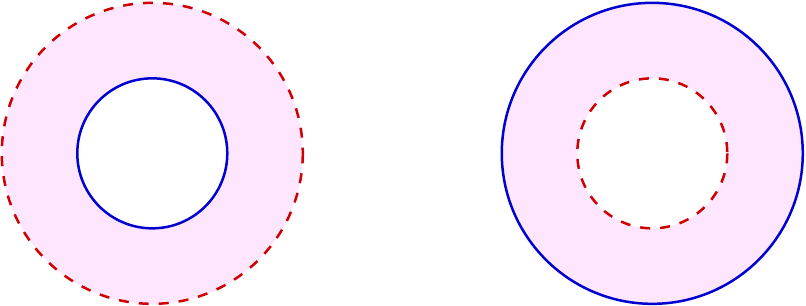}%
\end{picture}%
\setlength{\unitlength}{1579sp}%
\begingroup\makeatletter\ifx\SetFigFontNFSS\undefined%
\gdef\SetFigFontNFSS#1#2#3#4#5{%
  \reset@font\fontsize{#1}{#2pt}%
  \fontfamily{#3}\fontseries{#4}\fontshape{#5}%
  \selectfont}%
\fi\endgroup%
\begin{picture}(9658,3656)(362,-3179)
\put(5001,-1486){\makebox(0,0)[lb]{\smash{{\SetFigFontNFSS{10}{12.0}{\familydefault}{\mddefault}{\updefault}$+$}}}}
\end{picture}%
 \end{center}
 \caption{The one-loop diagrams with different conditions on the two boundaries.
Both orientations, exchanged by 
 $\Omega$, contribute. The solid line indicates the D7 boundary.}
 \label{fig:mixed_annuli}
\end{figure}
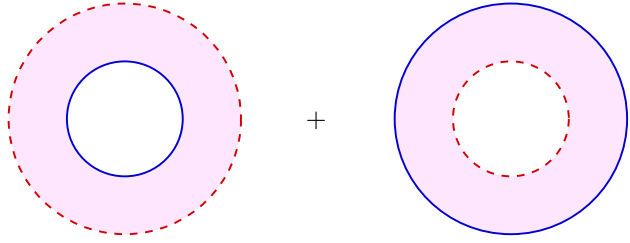

Let us now consider the contributions to the one-loop vacuum energy of the open strings stretching
between the $N$ D7 branes and the $k$ D-instantons, located at the same orientifold fixed
point%
\footnote{The complete and fully consistent expression of the instantonic vacuum amplitude should include also the annuli with one boundary attached to the D7 branes located at the other fixed points. These are important for the infrared properties in the closed string channel, but do not affect the infrared divergence in the open string one.}.
These contributions correspond to the mixed annuli diagrams
represented in Fig. \ref{fig:mixed_annuli}.

In such diagrams the open strings have mixed ND boundary conditions 
along the eight non-compact directions and DD boundary conditions along the two compact ones. 
For one string orientation the vacuum energy has the following schematic
form
\begin{equation}
Z_{\mathrm{D7/D(-1)}} = \int_0^\infty \frac{dt}{2t} \,\,
\mathrm{Tr}_{\mathrm{D7/D(-1)}}\left(\frac{1 + \Omega}{2}\,P_{\mathrm{GSO}}\, q^{L_0}\right)~.
\label{z7-1}
\end{equation}
Since $\Omega$ maps a state of the 7/(--1) sector into one of the (--1)/7 sector, it does not
contribute to the trace in (\ref{z7-1}), and we left with just one half of the annulus contribution.
In this amplitude the odd spin structure must be carefully evaluated, along the lines of \cite{Billo:2007sw,Billo:2007py} to which again we refer for details, and we find 
\begin{equation}
  Z_{\mathrm{D7/D(-1)}} = \frac{Nk}{2}
\int\frac{dt}{2t}\,\left\{\frac{1}{2}
\left[\Big(\frac{\theta_2(0|\ii t)}{\theta_4(0|\ii t)}\Big)^4
-\Big(\frac{\theta_3(0|\ii t)}{\theta_4(0|\ii t)}\Big)^4
-1
\right]\mathcal W(t)\right\}
\label{A71I}
\end{equation}
The three terms inside the square brackets correspond respectively to the NS, R and R$(-1)^F$
sectors, while there is no contribution from the NS$(-1)^F$ sector.
For the other orientation, we find instead
\begin{equation}
  Z_{\mathrm{D(-1)/D7}} = \frac{Nk}{2}
\int\frac{dt}{2t}\,\left\{\frac{1}{2}
\left[\Big(\frac{\theta_2(0|\ii t)}{\theta_4(0|\ii t)}\Big)^4
-\Big(\frac{\theta_3(0|\ii t)}{\theta_4(0|\ii t)}\Big)^4
+1
\right]\mathcal W(t)\right\}
\label{A17I}
\end{equation}
which is zero due to the abstruse identity of the $\theta$-functions.
Adding the two amplitudes (\ref{A71I}) and (\ref{A17I}), and using again the abstruse identity,
we find
\begin{equation}
Z_{\mathrm{D7/D(-1)}}+Z_{\mathrm{D(-1)/D7}} =
-\frac{Nk}{2}\int\frac{dt}{2t} \,\mathcal W(t)~.
\label{A71tot}
\end{equation}

Summing the various contributions, we easily find that the total one-loop vacuum
energy (\ref{oneloop}) is given by
\begin{equation}
\Gamma_{\mathrm{D(-1)}}=\frac{k(8-N)}{2}\int\frac{dt}{2t} \,\mathcal W(t)
\label{gamma1}
\end{equation}
which vanishes for $N=8$. This result is in perfect 
agreement with the calculation of the scaling dimension
of the measure on the instanton moduli space of the D7/D(--1) system, presented in 
Section \ref{subsec:scaling}. Furthermore, from (\ref{gamma1}) we can read that the one-loop action
of the brane system in the background of a single D-instanton is
\begin{equation}
 \label{Sinst1loop1}
S_{\mathrm{D(-1)}}^{\mathrm{1-loop}} \equiv -\Gamma_{\mathrm{D(-1)}}\Big|_{k=1} = \frac{N-8}2\int_0^\infty \frac{dt}{2t}\,\mathcal{W}(t)~.
\end{equation}
This is to be regarded as the one-loop correction to the tree-level classical instanton action
(\ref{smod1}) at $C_0=0$, which in the sequel we will denote by $S_{\mathrm{D(-1)}}^{\mathrm{tree}}$.

\subsection{Background fields vs D-instantons}
\label{subsec:bckg}
We now consider the D7-branes in another gauge configuration which is tractable at the string level, namely an abelian constant background. To remain general, we consider here an $\mathrm{SO}(N)$ configuration, even if in the specific applications we have in mind we set $N=8$ for local tadpole cancellation, as discussed in Section \ref{sec:system}.  We thus turn on a constant magnetic
field along, say, the directions $2$ and $3$, and take
\begin{equation}
\label{F23}
F_{23} = - F_{32} \equiv \mathcal{F}~,~~~~~
F_{\mu\nu}= 0~~\mbox{for~~$\mu,\nu\not=2,3$}~.
\end{equation}
In color space, this field can be diagonalized in the form
\begin{equation}
\label{fluxdef}
\mathcal F = \frac{1}{2\pi\alpha'}\, \mathrm{diag} \big(\ii f_1,\ii f_2, \ldots\big)~,
\end{equation}
with the eigenvalues $f_i$ ($i=1,\ldots N$) being paired into couples of opposite value. 
We have therefore
\begin{equation}
 \label{tracesF}
\begin{aligned}
& \Tr \mathcal F  = \frac{\ii}{2\pi\alpha'}\sum_i f_i = 0~,
& \Tr \mathcal F^2 = -\frac{1}{(2\pi\alpha')^2} \sum_i f_i^2~,\\
& \Tr \mathcal F^3  = - \frac{\ii}{(2\pi\alpha')^3}\sum_i f_i^3 = 0~,~~~
& \Tr \mathcal F^4  = \frac{1}{(2\pi\alpha')^4}\sum_i f_i^4~.
\end{aligned}
\end{equation}
For such a configuration, the expression (\ref{f4}) simplifies to
\begin{equation}
 \label{t8F23}
\mathrm{Tr}\big(t_8 F^4\big) = \frac 32\, \Tr \mathcal F^4~,
\end{equation}
while the topological term $(F\wedge F\wedge F\wedge F)$ vanishes, so that
the quartic action (\ref{S4nuova}) reduces to
\begin{equation}
 \label{actionF23}
S_{(4)}^{\mathrm{tree}}(\mathcal F) = -\frac{V_8}{16\lambda^4}\,\Tr \mathcal F^4
=-\frac{V_8}{64\pi^3 g_s}\,\Tr \mathcal F^4
\end{equation}
where $V_8$ is the (regularized) world-volume of the D7-branes. Recalling the expression of the
classical instanton action,
we can write the following relation
\begin{equation}
 \label{reldisk}
S_{(4)}^{\mathrm{tree}}(\mathcal F) = 
-\frac{V_8}{128\pi^4}\,\,\Tr \mathcal F^4\, \,S_{\mathrm{D(-1)}}^{\mathrm{tree}}~.
\end{equation}

In the following we will extend such a relation at the one-loop level. To this end
we first compute the one-loop vacuum energy of $N$ D7-branes of Type I$^\prime$
in the background (\ref{fluxdef}). This vacuum energy is schematically given by
\begin{equation}
 \label{1loopFdef0}
Z_{\mathrm{D7}}(\mathcal F) = \int_0^\infty \frac{dt}{2t}\, 
\mathrm{Tr}_{\mathcal F}\left(\frac{1 + \Omega}{2}\,P_{\mathrm{GSO}}\, q^{L_0}\right)
\end{equation} 
where the trace $\mathrm{Tr}_{\mathcal F}$ is computed over the space of all open string states 
attached to the D7-branes in the presence of the field $\mathcal F$.
Such a space separates into $N^2$ sectors with boundary conditions at $\sigma=(0,\pi)$ 
determined by the values $(f_i,f_j)$. These boundary conditions are of NN type 
in the directions 1,4,5,6,7,8, and of DD type in the directions 9,10, while in the complex direction 
$x = (x^2 + \ii x^3)/\sqrt{2}$ where there is the background field 
the string coordinate $x(z,\bar z)$ obeys this twisted boundary conditions
\begin{equation}
 \label{bc}
\partial x |_{\sigma=0} = \ee^{2\pi\ii \nu_i} \bar\partial x |_{\sigma=0}\quad,\quad
\partial x |_{\sigma=\pi} = \ee^{2\pi\ii \nu_j} \bar\partial x |_{\sigma=\pi}~,
\end{equation}
where $\pi\nu_{i,j} = -\arctan f_{i,j}$.
The field $x(z,\bar z)$ can be expressed as 
\begin{equation}
 \label{xtoX}
x(z,\bar z) = x_0 + \frac 12 \left[X(z) + \ee^{2\pi\ii \nu_i} X(\bar z)\right]
\end{equation}
in terms of a twisted chiral bosonic field $X(z)$ satisfying
\begin{equation}
 \label{Xtwist}
X(\ee^{2\pi\ii}z) = \ee^{2\pi\ii \nu_{ij}} X(z)\quad,\quad\mbox{with}~~ 
\nu_{ij} = \nu_i - \nu_j~.
\end{equation}
Similarly we can treat the corresponding world-sheet fermionic field $\psi = (\psi^2 + \ii \psi^3)/\sqrt{2}$, in its NS or R sector (see, for instance, \cite{Bertolini:2005qh} for
details).

Since world-sheet parity acts by exchanging the boundary conditions, according to
\begin{equation}
 \label {Omesunu}
 \Omega~~:~~ (\nu_i,\nu_j)~\to~ (-\nu_j,-\nu_i)~,
\end{equation}
the term in (\ref{1loopFdef0}) with the $\Omega$ insertion
contributes to the trace only when $\nu_j=-\nu_i$, that is 
when $f_j = -f_i$. The free energy $Z(\mathcal F)$ can thus be rewritten as
\begin{equation}
 \label{1loopFdef}
Z_{\mathrm{D7}}(\mathcal F) = \frac 12 \Bigg(\sum_{i,j} \mathcal{A}(f_i,f_j) 
+ \sum_{i} \mathcal{M}(f_i,-f_i)\Bigg)~,
\end{equation}
having denoted by $\mathcal{A}(f_i,f_j)$ and $\mathcal{M}(f_i,-f_i)$, respectively,
the annulus and the M\"obius diagrams in a specific sector.

The calculation of the annulus amplitudes in an external field is completely standard
(see for instance \cite{DiVecchia:2005vm}) and its explicit expression is 
\begin{eqnarray}
\mathcal{A}(f_i,f_j) & \equiv &
\int_0^\infty \frac{dt}{2t} \,
\mathrm{Tr}_{(f_i,f_j)}\left(P_{\mathrm{GSO}}\, q^{L_0}\right)
\label{Aij} \\
& = & \frac{\ii\,V_8}{(8\pi^2\alpha')^3}\frac{f_i - f_j}{4\pi^2\alpha'}
\int_0^\infty \frac{dt}{2t^4} \left[\frac{1}{2}\sum_{a=2}^4 s_a \,
\frac{\theta_a(\ii\nu_{ij}t|\ii t)}{\theta_1(\ii\nu_{ij}t|\ii t)} 
\Big(\!\!-\!2\pi\frac{\theta_a(0|\ii t)}{\theta_1'(0|\ii t)}\Big)^3
\right] \mathcal{W}(t)~,
\nonumber
\end{eqnarray}
with $a=2,3,4$ corresponding, respectively, to the R, NS and NS$(-1)^F$ sectors 
(hence $s_2 = -1$, $s_3=1$ and $s_4 =-1$).
In the above expression, the $\theta$-functions represent the contribution of the non-zero
modes of the various string fields; on the other hand a factor of $1/(8\pi^2\alpha' t)^3$ 
arises from the momentum integration over the six NN directions, while the non-commutativity of the
zero-modes in the twisted directions leads in the end to a factor of
$\ii (f_i-f_j)/(4\pi^2\alpha')$. The last term $\mathcal{W}(t)$ is as in
(\ref{W}) and represents the contribution of the winding modes along the two compact DD directions.

We are interested in the expansion of $\mathcal{A}(f_i,f_j)$ up to quartic order in the $f$'s.
To obtain it we first expand the even $\theta$-functions appearing in (\ref{Aij}) in powers
of $\nu_{ij}$, namely
\begin{equation}
 \theta_a(\ii\nu_{ij}t|\ii t)=\theta_a(0|\ii t) - 
\frac{1}{2}\,\nu_{ij}^2\tau^2\theta_a^{(2)}(0|\ii t) 
+ \frac{1}{4!}\,
\nu_{ij}^4t^4\theta_a^{(4)}(0|\ii t)+\cdots
\label{thetaexp}
\end{equation}
where $\theta_a^{(n)}$ denotes the $n$-repeated derivative of $\theta_a$ with respect to its first argument. When we substitute this expansion into (\ref{Aij}), we immediately see that the terms proportional to $\theta_a$ give rise to an expression proportional to the abstruse identity
\begin{equation}
 \label{aisa}
\sum_{a=2}^4 s_a \,\theta_a(0|\ii t)^4 = 0~;
\end{equation}
we can therefore forget about them. Similarly, the terms proportional to $\theta_a^{(2)}$ enter in (\ref{Aij}) only through a combination
which vanishes identically due to the following $\theta$-function 
identity (see Appendix \ref{app:theta}):
\begin{equation}
\label{idtheta2}
 \sum_{a=2}^4 s_a \,\theta_a^{(2)}(0|\ii t)\,\theta_a(0|\ii t)^3  = 0~.
\end{equation}
We are then left with only the terms involving $\theta_a^{(4)}$, which are already of quartic
order in the $f$'s since $\nu_{ij}=-(f_i-f_j)/\pi+O(f^3)$. This means that
in (\ref{Aij}) we can make the following replacement
\begin{equation}
\ii\,(f_i - f_j)\, \frac{\theta_a(\ii\nu_{ij}t|\ii t)}{\theta_1(\ii\nu_{ij}t|\ii t)}
\quad \to \quad-\,\frac{t^3(f_i-f_j)^4}{4!\,\pi^3}\,
\frac{\theta^{(4)}_a(0|\ii t)}{\theta'_1(0|\ii t)} + O(f^5)~,
\label{replacement}
\end{equation}
thus getting, up to quartic order in $f$,
\begin{equation}
 \label{Aij2}
\mathcal{A}(f_i,f_j) = \frac{V_8}{128 \pi^4} \left(\frac{f_i - f_j}{2\pi\alpha'}\right)^4
\int_0^\infty \frac{dt}{2t}\,\left[
\frac 16 \sum_{a=2}^4 s_a 
\frac{\theta_a^{(4)}(0|\ii t)\,\theta_a(0|\ii t)^3}{\theta'_1(0|\ii t)^4}
\right]
\mathcal{W}(t)~.
\end{equation}
Using the $\theta$-function identity, shown in Appendix \ref{app:theta},
\begin{equation}
 \label{idtheta4}
\sum_{a=2}^4 s_a \,\theta_a^{(4)}(0|\ii t)\,\theta_a(0|\ii t)^3
=3\,\theta'_1(0|\ii t)^4~,
\end{equation}
we see that the quantity in square brackets reduces to 1/2, so that we finally have
\begin{equation}
 \label{Aij3}
\mathcal{A}(f_i,f_j) = \frac{V_8}{256\pi^4} \left(\frac{f_i - f_j}{2\pi\alpha'}\right)^4 
\int_0^\infty \frac{dt}{2t}\,\mathcal{W}(t)~. 
\end{equation}
The sum of annuli amplitudes appearing in (\ref{1loopFdef}) can then be rewritten in terms 
of the traces of the matrix $\mathcal F$ using (\ref{tracesF}), getting
\begin{equation}
 \label{Aij4}
\sum_{i,j}\mathcal{A}(f_i,f_j) = \frac{V_8}{128\pi^4} \Big[N \,
{\mathrm{Tr}}\mathcal F^4 + 3 \big(\mathrm{Tr}\mathcal F^2\big)^2\Big]
\int_0^\infty \frac{dt}{2t}\,\mathcal{W}(t)~.
\end{equation}

Let us now consider the M\"obius diagrams $\mathcal{M}(f_i,-f_i)$. The action of $\Omega$ on the oscillators of the string which has $\nu_j=-\nu_i$ and twist $2\nu_i$ is the same as 
in the standard NN directions. As a result, the insertion of $\Omega$ has the net effect of sending $q=\ee^{-2\pi t} \,\to\, -q$, or equivalently  $\ii t\to \ii t +\frac12$, 
so that from the annulus amplitude (\ref{Aij}) we can immediately obtain the corresponding
M\"obius expression, namely 
\begin{eqnarray}
&& \mathcal{M}(f_i,-f_i)
\equiv 
\int_0^\infty \frac{dt}{2t} \,
\mathrm{Tr}_{(f_i,-f_i)}\left(\Omega\,P_{\mathrm{GSO}}\, q^{L_0}\right)
 \label{Mi}\\
&& = - \frac{\ii\,V_8}{(8\pi^2\alpha')^3}\frac{2f_i}{4\pi^2\alpha'}
\int_0^\infty \frac{dt}{2t^4} \left[\frac12\sum_{a=2}^4 s_a \,
\frac{\theta_a(2\ii\nu_{i}t|\ii t+\frac12)}{\theta_1(2\ii\nu_{i}t|\ii t
+\frac12)} \Big(\!\!-\!2\pi\frac{\theta_a(0|\ii t+\frac12)}{\theta_1'(0|\ii t+\frac12)}\Big)^3
\right] \mathcal{W}(t)
\nonumber
\end{eqnarray}
where the overall sign has been chosen to be consistent with the action of $\Omega$ on the Chan-Paton
factors of the 7/7 strings as given in (\ref{gammaomega7}).
The shift $\ii t \to \ii t + 1/2$ in the second argument of the $\theta$-functions
does not affect in any way the manipulations we performed in evaluating
the annulus amplitudes: indeed, both the expansion in powers of $f$ and the $\theta$-function identities to be used refer only to their first arguments. We can therefore immediately write, up to $O(f^5)$, 
\begin{equation} 
\label{resmoeb}
\mathcal{M}(f_i,-f_i)  = -\frac{V_8}{256\pi^4} \left(\frac{2f_i}{2\pi\alpha'}\right)^4 
\int_0^\infty \frac{dt}{2t}\,\mathcal{W}(t)~.
\end{equation} 
Summing over all M\"obius diagrams and using the trace formulas (\ref{tracesF}) gives us
\begin{equation} 
\label{resmoebtot}
\sum_i\mathcal{M}(f_i,-f_i)  = -\frac{V_8}{16\pi^4} \,{\mathrm{Tr}} \mathcal F^4
\int_0^\infty \frac{dt}{2t}\,\mathcal{W}(t)~.
\end{equation} 

Inserting the annulus and the M\"obius amplitudes (\ref{Aij3}) and (\ref{resmoebtot}) into  
(\ref{1loopFdef}), we have 
\begin{equation}
 \label{Fres}
Z_{\mathrm{D7}}(\mathcal F) = \frac{V_8}{256 \pi^4} \Big[(N-8) \,
{\mathrm{Tr}}\mathcal F^4 + 3 \big(\mathrm{Tr}\mathcal F^2\big)^2\Big]
\int_0^\infty \frac{dt}{2t}\,\mathcal{W}(t)+ O(f^5)~.
\end{equation}
We get therefore the following one-loop contribution to the single trace quartic action
\begin{equation}
 \label{s41loop}
S_{(4)}^{\mathrm{1-loop}}(\mathcal F) \equiv - Z_{\mathrm{D7}}(\mathcal F)\Big|_{\mathrm{quartic}}=
-\frac{V_8}{256 \pi^4}\, (N-8)\,{\mathrm{Tr}}\mathcal F^4 \int_0^\infty 
\frac{dt}{2t}\,\mathcal{W}(\tau)
~. 
\end{equation}
Comparing this expression to the one-loop action of the D7-branes in the background of a single D-instanton derived in the previous section (see Eq. (\ref{Sinst1loop1})), 
we obtain
\begin{equation}
 \label{relloop}
S_{(4)}^{\mathrm{1-loop}}(\mathcal F) = 
-\frac{V_8}{128\pi^4}\,\,\Tr \mathcal F^4\, \,S_{\mathrm{D(-1)}}^{\mathrm{1-loop}}~,
\end{equation}
which is the one-loop extension of the tree-level result written in (\ref{reldisk}).
What we have found is a generalization to the quartic term of the relation between 
instantonic one-loop amplitudes and gauge threshold corrections to the quadratic action, 
originally found in \cite{Abel:2006yk,Akerblom:2006hx}
and further elaborated in \cite{Billo:2007sw,Billo:2007py}.
The relations (\ref{reldisk}) and (\ref{relloop}) 
express the equality of the quartic coupling $\lambda$ computed in two different backgrounds: the
constant background $\mathcal F$ and the D(--1)-background. Furthermore, adding the one-loop contribution (\ref{s41loop}) to the corresponding tree-level expression (\ref{actionF23}) 
amounts to renormalize the coupling of the quartic action (\ref{S4nuova}) according to 
\begin{equation}
 \label{renquarticF}
 \frac{1}{\lambda^4} \to \frac{1}{\lambda^4}+ \frac{N-8}{16\pi^4}\, \int_0^\infty 
\frac{dt}{2t}\,\mathcal{W}(t)~.
\end{equation}
The integral appearing in the one-loop contribution is divergent both in the IR region, $t\to \infty$, and in the UV one, $t\to 0$.
It can be treated as described, for instance, in Appendix A of \cite{Billo:2007sw}, along the lines indicated in \cite{Dixon:1990pc}. The divergence for $t\to 0$ is associated
to massless exchanges in the closed channel, reached upon Poisson resummation,
and is absent when tadpole cancellation holds (which is actually the case when
$N=8$). Here, we are interested 
in the IR divergence, due to the massless open string states circulating in the loop, which we regularize in terms of a mass scale $\mu$ obtaining 
\begin{equation}
 \label{intW}
\int_0^\infty \frac{dt}{2t}\,\mathcal{W}(t)\to -\frac 12\log (\alpha'\mu^2) + \cdots
\end{equation}
Eq. (\ref{renquarticF}) can therefore be interpreted as a definition of the running coupling $\lambda(\mu)$:
\begin{equation}
 \label{lambdarunning}
 \lambda(\mu)^4 = \frac{\lambda^4}{1 +\frac{8-N}{32\pi^4}\,\lambda^4\,\log (\alpha'\mu^2)}~,
\end{equation}
with the original coupling $\lambda$ being defined at the scale $1/\sqrt{\alpha'}$. The one-loop $\beta$-function corresponding to (\ref{lambdarunning}) reads
\begin{equation}
 \label{betalambda}
\beta_\lambda \equiv \frac{\partial\log\lambda(\mu)}{\partial\log\mu} = -\frac{8-N}{64\pi^4}\lambda(\mu)^4 
\end{equation}
and we can introduce a renormalization group invariant scale $\Lambda$ at which the running coupling diverges given by
\begin{equation}
 \label{defLambda}
  \Lambda = \frac{1}{\sqrt{2\pi\alpha'}}\,\ee^{-\frac{16\pi^4}{(8-N)\lambda^4}} = 
\ell_s^{-1}\,\ee^{-\frac{4\pi}{(8-N)g_s}}~.
\end{equation}
Thus we can conclude that the prefactor (\ref{prefactor}) appearing in the instanton calculus
with D7 and D(--1)-branes can be expressed in terms of this renormalization group invariant
scale as
\begin{equation}
\Lambda^{\frac{8-N}{2}k}~,
 \label{prefactor1}
\end{equation}
in strict analogy with the usual instanton calculus with D3 and D(--1)-branes.
Clearly for $N=8$ all dimensional factors drop out and one is left with only the
classical factor $\ee^{-\frac{2\pi k}{g_s}}$ (or $\ee^{2\pi\ii\tau k}$ when $C_0\not=0$), 
similarly to what happens for the instanton contributions in $\mathcal N=4$ SYM theory.

In conclusion, we have shown that the one-loop renormalization of the quartic coupling $\lambda$
can be computed alternatively either utilizing a constant background field or assuming that the D-instanton represents the string realization of a classical configuration for which the 
quartic action $S_{(4)}$ reduces to the instanton action $S_{\mathrm{D(-1)}}$. 

\section{Eight-dimensional instantons}
\label{sec:eight}
As discussed in Sections \ref{sec:system} and \ref{sec:7-1}, the D(--1)-branes in our system are
related to some instanton-like gauge configuration in eight dimensions that satisfies
the relation (\ref{t8Fepsilon8F}) between the quartic action and the fourth Chern number.
Here we briefly review the known solutions to this constraint and argue that the $\mathrm{SO}(8)$ 
instanton discovered in \cite{Grossman:1984pi,Grossman:1989bb} is
singled out for our purposes.

\subsection{Linear instantons}
\label{subsec:lin}
One possibility to impose the constraint (\ref{t8Fepsilon8F}) is to exploit the relation between
$t_8$, $\epsilon_8$ and the octonionic projectors $P_1^\pm$ and $P_2^\pm$ discussed in Appendices
\ref{appa} and \ref{app:t8}. In particular, consider the relation (see Eq. (\ref{t8three}))
\begin{equation}
\label{t810}
t_8^{\mu_1\cdots\cdots\mu_8} =  \frac{1}{2}\,
\epsilon_8^{\mu_1\cdots\cdots\mu_8}
-\big(P^+_2-P^+_1\big)^{\mu_1\mu_2}_{\nu_1\nu_2}\cdots
\big(P^+_2-P^+_1\big)^{\mu_7\mu_8}_{\nu_7\nu_8}\,\epsilon_8^{\nu_1\cdots\cdots\nu_8}~,
\end{equation}
and take a gauge field $F$ belonging to $\mathrm{Ker}(P^+_1)$, namely such that
\begin{equation}
\big(P^+_1\big)^{\mu_1\mu_2}_{\nu_1\nu_2}\,F_{\mu_1\mu_2}=0\quad,\quad
\big(P^+_2\big)^{\mu_1\mu_2}_{\nu_1\nu_2}\,F_{\mu_1\mu_2}=F_{\nu_1\nu_2}~.
\label{FP1}
\end{equation}
Then, from (\ref{t810}) we obtain
\begin{equation}
t_8^{\mu_1\cdots\cdots\mu_8}\,F_{\mu_1\mu_2}\cdots F_{\mu_7\mu_8} = 
\frac{1}{2}\,\epsilon_8^{\mu_1\cdots\cdots\mu_8}\,F_{\mu_1\mu_2}\cdots F_{\mu_7\mu_8}  -
 \epsilon_8^{\nu_1\cdots\cdots\nu_8}\,F_{\nu_1\nu_2}\cdots F_{\nu_7\nu_8}~,
 \label{relFP1}
\end{equation}
from which the relation (\ref{t8Fepsilon8F}) immediately follows.
Using the explicit expressions of the projectors $P^+_1$ and $P^+_2$ given in (\ref{P12}),
we can rewrite the conditions (\ref{FP1}) as
\begin{equation}
F_{\mu\nu}+\frac{1}{2}\,C^{+}_{\mu\nu\rho\sigma}\,F^{\rho\sigma} = 0
\label{FC}
\end{equation}
where $C^+_{\mu\nu\rho\sigma}$ is the anti-symmetric four-index tensor constructed
from the octonionic structure constants. This is precisely the equation solved
by the Fubini-Nicolai instanton \cite{Fubini:1985jm}.
This is an instanton-like configuration with a $\mathrm{SO}(7)$ rotational symmetry and, if embedded in a supersymmetric context, it preserves 1/16 of supersymmetry as shown in \cite{Bak:2002aq}.
These, however, are not the symmetries of the D7/D(--1) system we are considering, which is
$\mathrm{SO}(8)$ invariant and 1/2-BPS. Thus, even if it satisfies the required
constraint (\ref{t8Fepsilon8F}), the Fubini-Nicolai instanton is not related to a D-instanton
inside a D7 brane.

It is worth to point out that Eq. (\ref{FC}) is just an example of more general
linear relations of the type
\begin{equation}
F_{\mu \nu} + \frac{1}{2}\,T_{\mu\nu\rho\sigma}F^{\rho\sigma}=0,
\label{linear}
\end{equation}
where $T$ is a constant anti-symmetric tensor, 
which were studied in \cite{Corrigan:1982th} as analogues in $d>4$ of the standard 
self-duality relation satisfied by the usual gauge instantons in $d=4$~\footnote{In $d=4$
the usual gauge (anti)instantons satisfy (\ref{linear}) with $T_{\mu\nu\rho\sigma}=\pm\epsilon_{\mu\nu\rho\sigma}$.}. These linear instantons
satisfy the Yang-Mills field equations $D^{\mu}F_{\mu\nu}=0$ as a consequence of the
Bianchi identity, and in general they may be classified by the unbroken rotational symmetry 
and the possibly unbroken supersymmetry. As shown in \cite{Bak:2002aq}, besides the
$\mathrm{SO}(7)$-invariant Fubini-Nicolai instanton we have mentioned above, in $d=8$ there exist
other linear instanton configurations that satisfy the constraint (\ref{t8Fepsilon8F}): 
these configurations are $\nu$-BPS, with $\nu=N/16$ for $N=2,\ldots,6$,
and are invariant under $\mathrm{SO}(N) \otimes \mathrm{SO}(8-N)$.
Also in these cases, the unbroken (super)symmetries do not match those
of the D7/D(--1)-brane system.

Another possibility to solve the constraint (\ref{t8Fepsilon8F}) is to consider a gauge field
that belongs to $\mathrm{Ker}(P^+_2)$,
so that the relation (\ref{relFP1}) is still satisfied. This is again a linear instanton configuration
of the type (\ref{linear}) with $T_{\mu\nu\rho\sigma}=-\frac13\,C^+_{\mu\nu\rho\sigma}$.
This background breaks the $\mathrm{SO}(8)$ rotational symmetry and, according to the
arguments of \cite{Bak:2002aq}, it cannot be promoted to a
BPS configuration in a supersymmetric context.
Nevertheless, it is interesting to point out that in this case the 
4-form $(F\wedge F)$ is self-dual, {\it i.e.}
\begin{equation}
\big(F\wedge F\big)^{\mu_1\mu_2\mu_3\mu_4} = \frac{1}{4!}\,\epsilon_8^{\mu_1
\cdots\cdots\mu_8}\,
\big(F\wedge F\big)_{\mu_5\mu_6\mu_7\mu_8} 
= *\big(F \wedge F\big)^{\mu_1\mu_2\mu_3\mu_4}~.
\label{selfduality0}
\end{equation}
This observation suggests that in order to impose the constraint (\ref{t8Fepsilon8F}) one
may look for gauge field configurations that satisfy the self-duality relation (\ref{selfduality0}), which is another way to generalize to $d=8$ what happens for the usual gauge instantons
in $d=4$. This is what we will describe in the next subsection.

\subsection{The $\mathrm{SO}(8)$ instanton}
\label{subsec:GKS}

The so-called $\mathrm{SO}(8)$ instanton \cite{Grossman:1984pi, Grossman:1989bb} 
corresponds to the following field configuration in $d=8$ \footnote{For an anti-instanton one has to replace $\gamma_{\mu\nu}$
with $\bar\gamma_{\mu\nu}$.
In the following, for definiteness we will consider only the instanton case.}:
\begin{equation}
\big[A_{\mu}(x)\big]^{\alpha\beta} = \frac{(\gamma_{\mu\nu})^{\alpha\beta}\,x^{\nu}}{r^2+\rho^2} 
\label{ainst}
\end{equation}
where $\rho$ is an arbitrary parameter representing the instanton size and $r^2=x_\mu x^\mu$.
Notice that the anti-symmetric pair $\alpha\beta$ of spinor indices
label the adjoint representation of $\mathrm{SO}(8)$, but, for simplicity, they will be often
omitted in the sequel. Notice also the complete similarity between (\ref{ainst}) and the
SU(2) instanton in $d=4$ in the regular gauge: the only differences are in the
range of the space-time indices and in the group structure since in the
four-dimensional case one finds the chiral spinorial generators of
$\mathrm{SU}(2)\subset\mathrm{SO}(4)$ in place of those of $\mathrm{SO}(8)$. 
The field strength corresponding to (\ref{ainst}) is
\begin{equation}
 F_{\mu\nu}(x)= \partial_\mu A_\nu(x)-\partial_\nu A_\mu(x) - \big[A_\mu(x),A_\nu(x)\big] =
-\frac{2\,\rho^2}{(r^2+\rho^2)^2}\,\gamma_{\mu\nu}~,
\label{fmunu}
\end{equation}
and satisfies $\big(F\wedge F\big)= *\big(F\wedge F\big)$
as a consequence of the self-duality relation
of the $\gamma_{\mu\nu}$ matrices 
(see Eq. (\ref{rel00})). The covariant derivative of this field strength reads 
\begin{equation}
 \label{KepDF}
\begin{aligned}
D_\mu F_{\nu\rho}(x) &= \partial_\mu F_{\nu\rho}(x) - \big[A_\mu(x),F_{\nu\rho}(x)\big]\\
&=\frac{4\rho^2}{(r^2+\rho^2)^3}\left(
2  x_{\mu} \gamma_{\nu \rho} -  x_{\nu} \gamma_{\rho \mu} 
+  x_{\rho} \gamma_{\nu \mu} +  \delta_{\mu \nu } \gamma_{\rho \sigma }
x^{\sigma} - \delta_{\mu \rho } \gamma_{\nu \sigma } x^{\sigma}
\right)~,
\end{aligned}
\end{equation}
from which, in $d$ dimensions, it immediately follows  that
\begin{equation}
 \label{notYM}
D^\mu F_{\mu\nu}(x) = \frac{4(d-4)\rho^2}{(r^2 + \rho^2)^3} \,\gamma_{\mu\nu}x^\nu~.
\end{equation}
Thus, the eight-dimensional $\mathrm{SO}(8)$ instanton (\ref{fmunu}) is \emph{not} a solution of the 
Yang-Mills equations, in contrast to the linear instantons considered in the previous subsection.

In a supersymmetric context, the $\mathrm{SO}(8)$ instanton describes a 1/2-BPS configuration 
\cite{Minasian:2001ib}; in fact by evaluating the gaugino supersymmetry transformation
\begin{equation}
 \delta \Lambda = \frac{1}{2}\,F_{\mu \nu}\,\Gamma^{\mu \nu} \,\varepsilon 
\label{gaugino}
\end{equation}
in the background (\ref{fmunu}) and taking into account its chirality, one 
finds 
\begin{equation}
 \delta (\Lambda_\gamma)^{\alpha\beta} \propto
 \frac{\rho^2}{(r^2+\rho^2)^2}\,\big(\delta^\alpha_\gamma\,\varepsilon^\beta-
 \delta^\beta_\gamma\,\varepsilon^\alpha\big)\quad,\quad
\delta (\Lambda_{\dot\gamma})^{\alpha\beta} =0~. 
\label{gauginos}
\end{equation}
The supersymmetry transformations associated to anti-chiral parameters $\varepsilon^{\dot\alpha}$
are then unbroken while those associated to chiral parameters $\varepsilon^\alpha$ are broken.
The field configuration (\ref{fmunu}) has then the same (super)symmetries of the D7/D(--1)-brane
system of Type I$^\prime$ we have described in the previous sections, and is naturally singled
out for our purposes.

Notice that the expression (\ref{gaugino}) is the
supersymmetric variation for the usual Yang-Mills action and that it is
corrected when the quartic action is taken into account. However
on the $\mathrm{SO}(8)$ instanton background it is the full answer as
shown in \cite{Minasian:2001ib}.
This instanton solution was originally proposed in \cite{Grossman:1984pi, Grossman:1989bb}
as a classical configuration that minimizes
the quartic action $\int \!d^8x\, \mathrm{Tr}(F\wedge F)^2$. Later it was realized \cite{Duff:1990wu, Minasian:2001ib} that it also minimizes the quartic action $S_{(4)}$ given in (\ref{S4nuova})
or equivalently that it satisfies the constraint
(\ref{t8Fepsilon8F}). To see this, let us start from the relation 
(see Eq. (\ref{t88}))
\begin{equation}
t_8 = -\frac{1}{2}\,\epsilon_8 + t_+~,
 \label{t888}
\end{equation}
and exploit the properties of the $\gamma_{\mu\nu}$ matrices (see in particular Eq. (\ref{tgg}))
to show that 
\begin{equation}
t_{+}^{\mu_1\cdots\cdots \mu_8}\,F_{\mu_5 \mu_6}F_{\mu_7 \mu_8}=0~,
\label{tf3}
\end{equation}
which in turn implies that
\begin{equation}
 \label{tf3bis}
t_8 F^4= -\frac{1}{2}\,\epsilon_8 F^4~.
\end{equation}

Let us now evaluate the quartic action on this instanton configuration.
Working formally in $d$ dimensions, after some straightforward algebra we have
\begin{equation}
 \mathrm{Tr}\big(t_8 F^4\big) = \left(-\frac{2\rho^2}{(r^2+\rho^2)^2}\right)^4
\left(-\frac{1}{8}\,d(d-1)(d-4)(7d-11)\right)
\Big(2^{d/2-1}\Big)~.
\label{trt8f4}
\end{equation}
The three factors above arise, respectively, from the fourth power of the instanton form
factor, the algebra of the $\gamma_{\mu\nu}$ matrices, and the trace over the
gauge indices which, with the Ansatz (\ref{ainst}) correspond to those of a Weyl spinor in
$d$ dimensions. Using 
\begin{equation}
 I_d=\int d^dx\,\frac{\rho^8}{(r^2+\rho^2)^8} 
= \frac{\pi^{d/2}\,\Gamma\big(8-d/2\big)}{7!}\,\rho^{d-8}~,
\label{Id}
\end{equation}
and setting $d=8$, we finally obtain
\begin{equation}
 -\frac{1}{4!\lambda^4} \int d^8x\, \mathrm{Tr}\big(t_8 F^4\big)= \frac{2\pi}{g_s}~,
\label{action4}
\end{equation}
which together with (\ref{t8Fepsilon8F}) implies that $c_{(4)}=1$. 
We can thus conclude that for the $\mathrm{SO}(8)$ instanton
\begin{equation}
 S_{(4)}= S_{\mathrm{D(-1)}}~.
\label{s4-1}
\end{equation}

Let us now briefly discuss the r\^ole of the fermionic sector of this instanton configuration.
As one can see from the supersymmetry transformations (\ref{gauginos}),
there is a non-trivial profile only for the chiral part of the gaugino 
and this is given by
\begin{equation}
(\Lambda_\gamma)^{\alpha\beta} =
 \frac{\rho^2}{(r^2+\rho^2)^2}\,\big(\delta^\alpha_\gamma\,\eta^\beta-
 \delta^\beta_\gamma\,\eta^\alpha\big)
 \label{gauginos1}
\end{equation}
where $\eta$ is an arbitrarily normalized fermionic parameter carrying a chiral spinor
index which can be taken to label the fundamental representation of $\mathrm{SO}(8)$.
We then consider the fermionic terms which supersymmetrize the quartic action $S_{(4)}$. 
Among them, the relevant one for our purposes is
\begin{equation}
\ii\,\frac{c}{\lambda^4}\,\int d^8 x \,
\mathrm{Tr}\big(F_{\mu\nu} F^{\mu\nu}\Lambda\,\big[\phi,\Lambda\big] \big)~,
\label{s4fer}
\end{equation}
where the complex scalar $\phi$ is the lowest component of the chiral superfield
$\Phi$ defined in (\ref{Phi}) and $c$ is a numerical coefficient. 
We observe that (\ref{s4fer}) descends from the term $\mathrm{Tr}\big(F_{MN} F^{MN} 
\chi \,\Gamma^P D_P \chi\big)$ in the NABI action in $d=10$, 
which is usually not written since it is proportional to the Dirac equation of motion.
However, we know that this term is present and has a non-vanishing coefficient $c$ since it is
found in the abelian Born-Infeld action. Despite the fact that the ten-dimensional
term is vanishing on-shell, it can give a non-zero contribution in $d=8$ 
if one considers a constrained instanton 
for which the Dirac equation (or its dimensionally reduced counterpart) does not hold.
This is indeed what happens when we take the $\mathrm{SO}(8)$ instanton in presence
of a non-vanishing vacuum expectation value for $\phi$, {\it i.e.} 
$\langle \phi^{\alpha\beta}\rangle=\phi_0^{\alpha\beta}$.
If we now insert the gauge field and gaugino profiles (\ref{fmunu}) and (\ref{gauginos1}) in the action (\ref{s4fer}), we get, up to a numerical factor which we absorb in a redefinition of $c$,
\begin{equation}
\ii\frac{c}{\lambda^4}\,{}^{\mathrm{t}}\eta\,\phi_0\,\eta\,\int d^8x\,\frac{\rho^8}{(r^2+\rho^2)^8} ~.
 \label{s4ferm1}
\end{equation}
Evaluating the integral using (\ref{Id}) for $d=8$ we see that all dependence on $\rho$ drops
out, and clumping again all numerical factors in $c$, we finally have
\begin{equation}
\ii\pi\frac{c}{g_s}\,{}^{\mathrm{t}}\eta\,\phi_0\,\eta ~.
 \label{s4ferm2}
\end{equation}
Writing $\eta=\sqrt{(g_s/c)}\,\mu'$ we can put this term in precisely the same form of
the fermionic part of the D-instanton moduli action derived in Section \ref{subsec:modact}.
We therefore conclude that the $\mathrm{SO}(8)$ superinstanton completely accounts for the D(--1) action $S_{\mathrm{D(-1)}}(\Phi)$ given in (\ref{smod12}), when the vacuum expectation
value $\phi_0$ is promoted to the full-fledged superfield $\Phi(x,\theta)$.

\section{An almost vacuum configuration}
\label{sec:vacuum}
The SO(8) instanton we described in the previous section has a free parameter $\rho$ corresponding to its size. In the four dimensional case, the size is one of the parameters encoded in the moduli $w_{\dot\alpha}$ of the ADHM construction.
As we have discussed in Section \ref{subsec:spectrum},
the D7/D(--1) string does not possess the vertex operators corresponding to
these moduli. In this section we consider the consequences of this absence on
the field configuration associated to the D(--1)-branes.

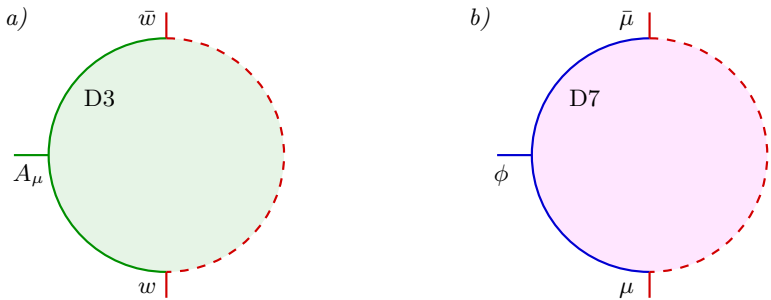
\begin{figure}[htb]
 \begin{center}
 \begin{picture}(0,0)%
\includegraphics{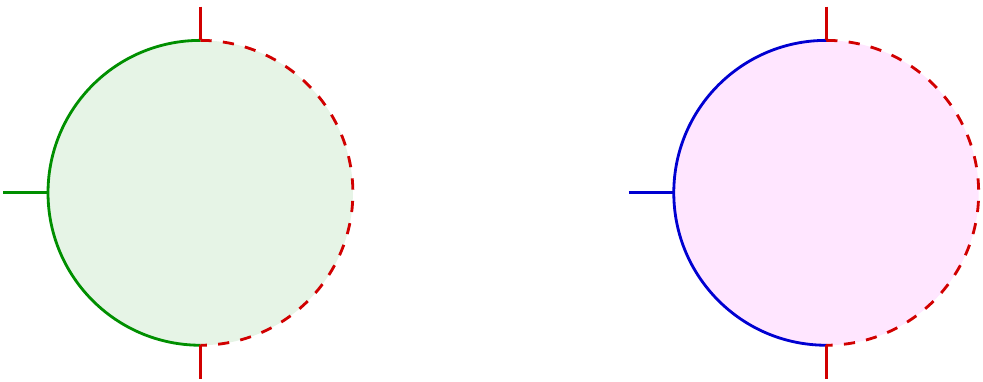}%
\end{picture}%
\setlength{\unitlength}{2052sp}%
\begingroup\makeatletter\ifx\SetFigFontNFSS\undefined%
\gdef\SetFigFontNFSS#1#2#3#4#5{%
  \reset@font\fontsize{#1}{#2pt}%
  \fontfamily{#3}\fontseries{#4}\fontshape{#5}%
  \selectfont}%
\fi\endgroup%
\begin{picture}(8523,3543)(136,-3073)
\put(7401,-2986){\makebox(0,0)[lb]{\smash{{\SetFigFontNFSS{9}{10.8}{\familydefault}{\mddefault}{\updefault}$\mu$}}}}
\put(7401,239){\makebox(0,0)[lb]{\smash{{\SetFigFontNFSS{9}{10.8}{\familydefault}{\mddefault}{\updefault}$\bar \mu$}}}}
\put(5901,-1636){\makebox(0,0)[lb]{\smash{{\SetFigFontNFSS{9}{10.8}{\familydefault}{\mddefault}{\updefault}$\phi$}}}}
\put(6801,-736){\makebox(0,0)[lb]{\smash{{\SetFigFontNFSS{9}{10.8}{\familydefault}{\mddefault}{\updefault}$\mathrm{D}7$}}}}
\put(5601,239){\makebox(0,0)[lb]{\smash{{\SetFigFontNFSS{9}{10.8}{\familydefault}{\mddefault}{\updefault}$\emph{b)}$}}}}
\put(1651,-2986){\makebox(0,0)[lb]{\smash{{\SetFigFontNFSS{9}{10.8}{\familydefault}{\mddefault}{\updefault}$w$}}}}
\put(1651,239){\makebox(0,0)[lb]{\smash{{\SetFigFontNFSS{9}{10.8}{\familydefault}{\mddefault}{\updefault}$\bar w$}}}}
\put(151,-1636){\makebox(0,0)[lb]{\smash{{\SetFigFontNFSS{9}{10.8}{\familydefault}{\mddefault}{\updefault}$A_\mu$}}}}
\put(1001,-736){\makebox(0,0)[lb]{\smash{{\SetFigFontNFSS{9}{10.8}{\familydefault}{\mddefault}{\updefault}$\mathrm{D}3$}}}}
\put(51,239){\makebox(0,0)[lb]{\smash{{\SetFigFontNFSS{9}{10.8}{\familydefault}{\mddefault}{\updefault}$\emph{a)}$}}}}
\end{picture}%
 \end{center}
 \caption{\emph{a)} The emission of the gauge field from a mixed disk in the D3/D(--1) system 
explains the classical profile of the instanton. \emph{b)} In the D7/D(--1) system,
no emission diagram for the gauge field is found, owing to the absence of the $w$ moduli.
There is only an emission diagram for the scalar $\phi$, involving the fermionic moduli $\mu$.}
 \label{fig:emissions}
\end{figure}

In the D3 case, the correspondence between D-instantons and gauge instantons is made very explicit by the fact that the D(--1)'s represent the source for the classical instanton solution \cite{Billo:2002hm}. From the diagram represented in Fig. \ref{fig:emissions}\emph{a)},
one gets (in the SU(2) case, for $c_{(2)}=1$)%
\footnote{We use the same notations adopted in 
\cite{Billo:2002hm}.}
\begin{equation}
\label{gf1}
\big[A_\mu(x)\big]^{uv}  =  w_{\dot\alpha}^{~u}\,
(\bar\sigma_{\mu\nu})^{\dot\alpha}_{~\dot\beta}\, \bar w^{\dot \beta\,v}
\,\frac{x^\nu}{r^4}  = \rho^2\,(\bar\sigma_{\mu\nu})^{uv} \,\frac{x^\nu}{r^4}~.
\end{equation}
In the second step above, using the ADHM constraints, the moduli dependence has been re-expressed in terms of the size $\rho$, which is related to the $w$ and $\bar w$ moduli by
\begin{equation}
  \label{rhodef}
 2 \rho^2 = \bar w^{\dot \alpha}_{~u} w_{\dot \alpha}^{~u}~.
\end{equation}

Eq. (\ref{gf1}) represents the leading behaviour for large distances, {\it i.e.} for
$r^2/\rho^2\to \infty$, of the instanton solution in the singular gauge
\begin{equation}
 \label{4dinstsing}
A_\mu(x) = \frac{\rho^2\,\bar\sigma_{\mu\nu}\,x^\nu}{r^2(r^2+\rho^2)}~.
\end{equation}
Subleading  orders of the large-distance expansion are provided by considering more and more source terms of this kind. 

In absence of the source represented by the diagram in Fig. \ref{fig:emissions}\emph{a)}, no classical instanton profile of the D3 gauge field would be associated to the D-instanton. 
This, however, is exactly the situation which occurs in the D7/D(--1) system:
there are no bosonic moduli $w$ and $\bar w$ 
from the NS sector of mixed D7/D(--1) strings, and no emission diagram for the gauge field 
like the one of Fig. \ref{fig:emissions}\emph{a)} can be constructed.
The only diagram describing the emission of a D7/D7 field is the one in Fig. \ref{fig:emissions}\emph{b)}: it acts as a source for  the scalar $\phi$ and contains only fermionic moduli. 

This simple fact seems to imply that the classical instantonic configuration of the gauge field on the D7 branes associated to a D(--1) \emph{vanishes}, except at the location of the D(--1) itself.
The only possibility is that the sought-for configuration is represented by the zero-size limit $\rho\to 0$ of the SO(8) instanton configuration we singled out in the previous section.

\section{The field theory limit and the vanishing of higher $\alpha'$ corrections}
\label{sec:vanishing}
To complete our analysis we now discuss the interplay between the zero-size limit $\rho\to 0$
of the $\mathrm{SO}(8)$ instanton configuration and the limit $\alpha'\to 0$.

At the string level, the effective action of the D7-branes is the full NABI action: this is organized
as a series of contributions with increasing powers of $\alpha'$, the first few of which are
given in Eq. (\ref{action0}) that we rewrite here for convenience:
\begin{equation}
 S_{\mathrm{D7}}  = \frac{1}{128\pi^5\alpha'^2\, g_s}\int d^8x \,\mathrm{Tr}
\big(F^2\big)
- \frac{1}{96\pi^3g_s}\int d^8x \,\mathrm{Tr}\big(t_8\,F^4\big) 
+ \frac{\alpha'}{g_s} \int d^8x\, \mathcal{L}_{(5)}(F,DF)+\cdots~.
\label{action00}
\end{equation}
In order to keep the quartic term and describe the effects of the D-instantons in the field theory limit, we should take the limit $\alpha'\to 0$ with $g_s$ fixed. 
This is dangerous because the quadratic Yang-Mills term in (\ref{action00}) 
naively explodes being proportional to
$1/(\alpha')^{2}$. However, it is easy to show that on a solution like the $\mathrm{SO}(8)$ instanton
which is localized within a region of size $\rho$, the Yang-Mills action vanishes in the zero-size
limit. Indeed, inserting the field strength (\ref{fmunu}) into $S_{\mathrm{YM}}$
and formally working in $d$ dimensions, we get
\begin{equation}
 \label{KepYMint}
\frac{2^{d/2-1}}{32\pi^5\alpha'^2\, g_s}\, d(d-1) \int d^dx \, \frac{\rho^4}{(r^2 + \rho^2)^4}~.
\end{equation}
For $d\to 8\,$ the integral is divergent and we have to regulate it. For instance,
using an explicit cut-off $R$, we get a result proportional to
\begin{equation}
\frac{\rho^4}{\alpha'^2\, g_s} \log\big(\rho/R\big)~,
\end{equation}
whereas if we use dimensional regularization setting $d= 8 - 2\epsilon$, up to numerical
factors we obtain 
\begin{equation}
\label{dimregYM}
\frac{\rho^4}{\alpha'^2\, g_s} \left(\frac{1}{\epsilon} + \mbox{finite}\right)~.
\end{equation}
In both cases, we clearly see that by taking
the limit $\rho\to 0$ before removing the regulator, the Yang-Mills contribution vanishes. 
It is interesting to note that the same thing happens also for the quadratic fermionic
terms which supersymmetrize the Yang-Mills action. In fact, by inserting 
in $\int d^8 x \,\mathrm{Tr}\big(\Lambda\,\big[\phi,\Lambda\big] \big)$
the gaugino profile (\ref{gauginos1}) and giving a vacuum expectation value
to the scalar field $\phi$, we obtain the same integral as in (\ref{KepYMint}),
so that also the quadratic fermionic contribution vanishes in the limit $\rho\to 0$.

This analysis of the Yang-Mills action shows that in order to associate a classical field configuration to the D7/D(--1)-brane system, one has first to take the limit $\rho\to 0$ of the $\mathrm{SO}(8)$ (super)instanton, and then take the limit $\alpha'\to 0$ (with $g_s$ fixed).
Of course, this limiting procedure has to be consistent also on the other terms of the NABI action.
As discussed in Section~\ref{sec:eight}, the quartic term does not create any problem in this respect
because it produces structures which depend neither on $\alpha'$ nor on $\rho$. On the other hand,
by simple dimensional analysis, the contributions to the NABI action (\ref{action00}) of higher
order in $\alpha'$, when calculated on the $\mathrm{SO}(8)$ instanton, give rise 
(in $d$ dimensions) to a series of the form
\begin{equation}
 \label{Kepseries}
\rho^{d-8}\,\sum_{n=1}^\infty a_n \, \left(\frac{\alpha'}{\rho^2}\right)^n~,
\end{equation}
where for instance the coefficient $a_1$ is determined by evaluating the integral of the quintic 
Lagrangian $\mathcal{L}^{(5)}(F,DF)$ on the instanton solution.
Since we have to take the point-like limit $\rho\to 0$ before the field-theory limit $\alpha'\to 0$, for the consistency of the whole construction it is necessary 
that the coefficients $a_n$ be zero, {\it i.e.} it is necessary that all higher order contributions
in the NABI action vanish when evaluated on the $\mathrm{SO}(8)$ solution.
However, as the NABI action is not known in a closed form, 
one can only check the first coefficients of the expansion (\ref{Kepseries}).

Before performing such checks, it is worth recalling a few facts about the structure of the NABI action. The quartic term in (\ref{action00}), which is of order $O(\alpha'{}^2)$ with respect to the Yang-Mills one, was computed long time ago by Tseytlin \cite{Tseytlin:1986ti} starting from the evaluation of 4-point string amplitudes. In this seminal paper it is clearly discussed how the identification of the effective action through on-shell amplitudes is inherently ambiguous with respect to the presence of terms proportional to $D^\mu F_{\mu\nu}$. The structure 
$\Tr \big(t_8 F^4\big)$ of the quartic Lagrangian corresponds to a ``minimal'' scheme in which terms of this type are absent, and can be related to other schemes by field redefinitions. This same ambiguity obviously persists in the subsequent higher-order terms.
A way to fix this ambiguity can be to insist that the bosonic NABI action should admit 
an off-shell supersymmetric extension; this indeed singles out the minimal form of the quartic Lagrangian $\Tr \big(t_8 F^4\big)$ as shown in \cite{Cederwall:2001bt,Cederwall:2001td,Bergshoeff:2001dc}.
This procedure has been extended in \cite{Collinucci:2002ac}
also to the terms of $O(\alpha'{}^3)$, {\it i.e.} to the quintic
Lagrangian $\mathcal{L}^{(5)}(F,DF)$ appearing in (\ref{action00}), which in this way
is determined to be 
\begin{eqnarray}
\label{L5}
\mathcal{L}^{(5)} & = &\frac{\zeta(3)}{2}\,
\Tr\Bigl\{ 4 \big[{F_{\mu_1 \mu_2}},{ F_{\mu_3 \mu_4}}\big] \Bigl[{\big[{F_{\mu_1
\mu_3}},{ F_{\mu_2 \mu_5}}}\big],{ F_{\mu_4 \mu_5}}\Bigr]\\
 && +\, 2 \big[{F_{\mu_1 \mu_2}},{ F_{\mu_3 \mu_4}}\big] \Bigl[{\big[{F_{\mu_1 \mu_2}},{
F_{\mu_3 \mu_5}}}\big],{ F_{\mu_4 \mu_5}}\Bigr]
+  2 \big[{F_{\mu_1 \mu_2}},{ D_{\mu_5} F_{\mu_1 \mu_4}}\big] \big[{D_{\mu_5}F_{\mu_2
\mu_3}},{F_{\mu_3 \mu_4}}\big] \nonumber\\
&&   -\,2 \big[{F_{\mu_1 \mu_2}},{ D_{\mu_4} F_{\mu_3 \mu_5}}\big] \big[{D_{\mu_4} F_{\mu_2
\mu_5}},{F_{\mu_1 \mu_3}}\big]
+   \big[{F_{\mu_1 \mu_2}},{ D_{\mu_5} F_{\mu_3 \mu_4}}\big] \big[{D_{\mu_5}
F_{\mu_1 \mu_2}},{F_{\mu_3 \mu_4}}\big]
\Bigr\}~.
\nonumber
\end{eqnarray}
It is natural to consider this quintic Lagrangian as the first string correction to the minimal quartic
Lagrangian $\Tr \big(t_8 F^4\big)$, since its structure is fixed using the same guiding principles.
For this reason the Lagrangian (\ref{L5}) is singled out among the various proposals for $\mathcal{L}^{(5)}$ existing in the literature.

Plugging the instanton profiles (\ref{fmunu}) and (\ref{KepDF}) into 
$\frac{\alpha'}{g_s} \int\! d^8x \,\mathcal{L}^{(5)}$ and working formally in $d$ dimensions,
after straightforward algebraic manipulations that we performed with the help of
the XCadabra program \cite{Peeters:2006kp,Peeters:2007wn}, from (\ref{L5}) we find
\begin{equation}
 \label{res1}
\begin{aligned}
\frac{\alpha'\,\zeta(3)}{g_s}\, 2^{d/2+9}
\Bigl(&- d(d-1)(d-2)(d-4) \int d^dx \,\frac{\rho^{10}}{(r^2 +
\rho^2)^{10}}\\
 &+  (d-1)(d-2)(d-4)(d+2)\int d^dx \,\frac{\rho^8\, r^2}{(r^2 +
\rho^2)^{10}}
\Bigr)~.
\end{aligned}
\end{equation}
The first line represents the contribution of the $F^5$ terms, while the second stems
from the $(DF)^2 F^2$ ones.
Performing the integrations, in the end we remain with
\begin{equation}
 \label{res2}
\frac{\alpha'\,\zeta(3)}{g_s}\,2^{d/2+9} \,
\frac{\pi^{d/2}\,\Gamma(9-d/2)}{9!\,\rho^{10-d}}
\,(d-1)(d-2)(d-4)\left(-d\Big(9-\frac d2\Big) + \big(d+2\big)\frac d2\right)~,
\end{equation}
so that the coefficient $a_1$ in (\ref{Kepseries}) takes the form
\begin{equation}
 \label{c1res}
a_1 \propto d(d-1)(d-2)(d-4)(d-8)~.
\end{equation}
We therefore see that the $O(\alpha'{})$ corrections vanish on the
instanton Ansatz (\ref{ainst}) for $d=8$. 
We believe that this is a very non-trivial check of the consistency 
of our picture in which the zero-size limit of the $\mathrm{SO}(8)$ instanton 
is identified as the field theoretical counterpart of the D7/D(--1)-brane system.

It is curious to notice that the coefficient $a_1$ vanishes for $d=1,2,4,8$ which are the dimensions
of the four division algebras over the reals, namely the real, complex, quaternionic and
octonionic algebras. These algebras are related to the four fundamental Hopf maps, which
physically describe, respectively, the kink solution of the Sine-Gordon equation, 
the Dirac monopole, the Yang-Mills gauge instantons in four dimensions 
and the SO(8) instanton in eight dimensions \cite{Grossman:1984pi,Grossman:1989bb}.

We remark that in the literature there are different
forms for the $O(\alpha'^3)$ Lagrangian $\mathcal{L}^{(5)}$, which have been determined by
fixing the field redefinition ambiguity mentioned above by focusing on different guiding principles. 
For instance, in \cite{Refolli:2001df,Grasso:2002wb} the Lagrangian is derived from superfield loop computations, while in \cite{Barreiro:2005hv} it is obtained from superstring amplitudes. In \cite{Koerber:2001uu}, instead, the requirement is that a particular class of BPS solutions of the Yang-Mills equations, called holomorphic instantons, remain solutions also of the NABI action. Such solutions correspond to the usual gauge instantons in $d=4$, but do not correspond to the $\mathrm{SO}(8)$ instanton in $d=8$ which is the relevant one for the D7/D(--1) system under consideration.
As stated in the literature, all proposed forms of $\mathcal{L}^{(5)}$ agree among each other up to terms proportional to the Yang-Mills field equations.
Indeed, when evaluated on the Ansatz (\ref{fmunu}), we find that the corresponding actions all vanish in $d=4$, where the Yang-Mills equations are satisfied; they however disagree in the $d=8$ case, where only the supersymmetrizable action corresponding to (\ref{L5}) vanishes. 

As mentioned above, we believe that the expression compatible with off-shell supersymmetry is the one that should be used in order to weight configurations such as the eight-dimensional 
instanton which is ``off-shell'' because it does not satisfy the Yang-Mills field
equations. Finally, we observe that the $O(\alpha'{}^4)$ correction to the NABI action has been derived in \cite{Koerber:2002zb}, extending the same philosophy
used in \cite{Koerber:2001uu} for the $O(\alpha'{}^3)$ term. We can therefore expect that it
coincides with the off-shell supersymmetrizable expression only up to terms proportional to
$D^\mu F_{\mu\nu}$ which are not negligible on our instanton solution. So, 
even if it would be highly desirable to be able to check the vanishing of the
action also to $O(\alpha'{}^4)$, and hence of the coefficient $a_2$ in
(\ref{Kepseries}), 
this does not seem to be possible at the moment. Nevertheless, we believe that the vanishing of
the first coefficient $a_1$ is already a highly non-trivial test of our
proposal.

\paragraph{Concluding remarks}
We think that the results and techniques of this paper can be further developed
in several directions. In particular, in the eight-dimensional context,
it would be obviously very interesting to perform explicitly the moduli integral
in (\ref{npea}) to determine the coefficients $c_k$ of (\ref{npea1}) and
compare them to those derived from the duality with the heterotic string on
$T_2$. Regarding our starting motivation, namely the relation with
so-called exotic instantons, one could repeat the analysis of this paper upon
further compactification on, say, a $T_4$. The resulting systems of wrapped D7's
and D(--1)'s  represent four-dimensional exotic instantons, albeit within a system with
$\mathcal{N}=4$ supersymmetry; it would be nice to understand if they possess
a field-theoretic interpretation related to a compactification of the zero-size
limit of the eight-dimensional SO(8) solution, and if some similar
interpretation can be found also in cases with reduced supersymmetry.

\vskip 1cm
\noindent {\large {\bf Acknowledgments}}
\vskip 0.2cm
\noindent We would like to especially thank C. Bachas and P. Di Vecchia for many discussions and exchange of ideas, and for sharing with us notes and calculations.
We would like to thank also F. Fucito and J. F. Morales for several illuminating discussions. 
M.B. and A.L. thank the E. Schr\"odinger Institute in Vienna for hospitality.
This work has been partially supported by the European Commission FP6 Programme under contract MRTN-CT-2004-005104 ``{Constituents, Fundamental Forces and Symmetries of the Universe}''.
\appendix

\section{Octonions and $\mathrm{SO}(8)$ $\Gamma$-matrices}
\label{appa}

\subsection{Octonions}
\label{app:octo}
An octonion can be defined as  $q = q^{\mu}\,e_{\mu}$,
where the eight components $q^{\mu}$ are real numbers, and the eight basis vectors are $e_8=1$
and $e_i$ ($i=1,2,...,7$), such that
\begin{equation}
e_i \,e_j = -\delta_{ij} + c_{ijk}\, e_k
\end{equation}
with $c$ a totally antisymmetric tensor whose only non-zero elements can be taken to be
\begin{equation}
c_{127}=c_{163}=c_{154}=c_{253}=c_{246}=c_{347}=c_{567}=1~.
\label{c}
\end{equation}
One can easily verify that the tensor $c$ obeys, among others, the following identities
\begin{equation}
 \begin{aligned}
  &c_{ijk}\,c_{k\ell m} = \delta_{i\ell}\,\delta_{jm} - \delta_{im}\,\delta_{j\ell} + 
  \frac{1}{3!}\, \epsilon_{ij\ell mnpq}\,c_{npq} ~,\\
  &c_{ijk}\,c_{jk\ell} = 3!\, \delta_{i\ell}~,\\
 &c_{ijk} = \frac{1}{4!}\,\epsilon_{ij\ell mnpq}\,c_{npq}\,c_{k\ell m}
~.
 \end{aligned}
\label{idC}
\end{equation}
The tensor $c$ can be embedded into two totally antisymmetric four-index tensors
$C_{\mu\nu\rho\sigma}^\pm$ in $d=8$ by means of
\begin{equation}
C_{ijk8}^\pm = c_{ijk} \quad,\quad
C_{ijk\ell}^\pm = \pm \frac{1}{3!} \,\epsilon_{ijk\ell mnp}\,c_{mnp}
\label{Cpm}
\end{equation}
{From} (\ref{c}) we see that, up to permutations of the indices, the only non-vanishing
components of $C^\pm$ are
\begin{equation}
 \begin{aligned}
 & C_{1278}^\pm=C_{1638}^\pm=C_{1548}^\pm=C_{2538}^\pm=C_{2468}^\pm=C_{3478}^\pm=C_{5678}^\pm = + 1 ~,
\\
& C_{1234}^\pm =C_{1256}^\pm=C_{1357}^\pm=C_{1647}^\pm=C_{3267}^\pm=C_{4257}^\pm=C_{3456}^\pm = \pm 1 ~.
 \end{aligned}
\label{Cpm1}
\end{equation}
The two tensors $C^\pm$ obey the following (anti)self-duality relations in $d=8$
\begin{equation}
C^{\pm\,\mu_1\mu_2\mu_3\mu_4}= 
\pm\,\frac{1}{4!}\,\epsilon_8^{\mu_1\cdots\cdots\mu_8}
\,C^{\pm}_{\mu_5\mu_6\mu_7\mu_8}
\label{selfduality}
\end{equation}
and thus transform respectively in the representations 
$\mathbf{35}$ and $\mathbf{35'}$ of $\mathrm{SO}(8)$.

The identities satisfied by $c$, like the ones in (\ref{idC}), can be re-expressed as
identities on $C^\pm$. In particular, we find useful to mention the following ones
\begin{equation}
 \begin{aligned}
  &\frac{1}{2}\,C^{\pm\, \mu\nu\lambda\tau} \,C^{\pm}_{\lambda\tau\rho\sigma } = 3 \,
  \delta^{\mu\nu}_{\rho\sigma} \pm 2 C^{\pm \,\mu\nu}_{\rho\sigma}~,\\
& \frac{1}{3!}\,C^{\pm\,\mu\rho\sigma\tau}\,C^{\pm}_{\nu\rho\sigma\tau} = 7\, \delta^{\mu}_{\nu}~,\\
& \frac{1}{4!}\,C^{\pm\mu\nu\rho\sigma}\,C^{\pm}_{\mu\nu\rho\sigma} = 14~,
 \end{aligned}
\label{idenC}
\end{equation}
where we have used the notation 
$\delta^{\mu\nu}_{\rho\sigma}=\delta^\mu_\rho\delta^\nu_\sigma - \delta^\nu_\rho\delta^\mu_\sigma$.
Following \cite{deWit:1983gs}, we can exploit the properties of $C^\pm$ to define the 
operators
\begin{equation}
\big(P^{\pm}_1\big)^{\mu\nu}_{\rho\sigma} 
= \frac{1}{8} \Big(\delta^{\mu\nu}_{\rho\sigma} \pm C^{\pm\,\mu\nu}_{\rho\sigma}\Big)\quad,\quad \big(P^{\pm}_2\big)^{\mu\nu}_{\rho\sigma} 
= \frac{3}{8} \Big(\delta^{\mu\nu}_{\rho\sigma} \mp\frac{1}{3}\, C^{\pm\,\mu\nu}_{\rho\sigma}\Big)
\label{P12}
\end{equation}
which act as orthogonal projectors on the 28-dimensional space of the $(8\times 8)$ anti-symmetric matrices. Indeed, using (\ref{idenC}) it is easy to check that
\begin{equation}
 \begin{aligned}
&P^\pm_1\cdot P^\pm_1 = P^\pm_1\quad,\quad P^\pm_2\cdot P^\pm_2 = P^\pm_2
~,\\
&P^\pm_1\cdot P^\pm_2 = P^\pm_2\cdot P^\pm_1 = 0\quad,\quad
P^\pm_1+P^\pm_2 = \frac12\,\delta~,
 \end{aligned}
\label{propP12}
\end{equation}
where we have used the notation $\big(A\cdot B\big)^{\mu\nu}_{\rho\sigma}
= A^{\mu\nu}_{\lambda\tau}\,B^{\lambda\tau}_{\rho\sigma}$, and understood all indices.
Thus, any anti-symmetric tensor of rank two can be decomposed into two independent pieces, 
one annihilated by $P^\pm_1$ and one annihilated by $P^\pm_2$. Since the $C^\pm$ tensors are
traceless, the dimensionalities of the two eigenspaces can be easily obtained
by computing the trace of the two projectors, and one finds
\begin{equation}
 \mathrm{dim}\big[\mathrm{Ker}(P^\pm_1)\big] = 21 \quad,\quad 
\mathrm{dim}\big[\mathrm{Ker}(P^\pm_2)\big] = 7~.
\label{dim}
\end{equation}

The octonion structure constants can be used also to construct an explicit realization of the Clifford algebra in $d=7$. Indeed, it is easy to check that 
the seven $(8\times 8)$-matrices $\tau^i$ ($i=1,\ldots,7$) with
elements
 \begin{equation}
(\tau^i)_{\alpha\beta} =
 \delta^{i8}_{\alpha\beta}+C^{-\,i8}_{\alpha\beta}\quad\quad(\alpha,\beta=1,\ldots, 8)
 \label{gamma7}
\end{equation}
satisfy the following relations
\begin{equation}
\begin{aligned}
&\big\{\tau^i,\tau^j\big\}_{\alpha\beta} = -2\delta^{ij}\,\delta_{\alpha\beta}~,\\
&(\tau^{ij})_{\alpha\beta} 
\equiv \frac{1}{2}\big[\tau^i,\tau^j\big]_{\alpha\beta} = -\delta^{ij}_{\alpha\beta}
-C^{-\,ij}_{\alpha\beta}~.
\end{aligned}
\label{rel}
\end{equation}
Furthermore, by direct computation or by using the properties of the $C^-$ tensor, one can show that
\begin{equation}
\tau^{1}\tau^{2}\tau^{3}\tau^{4}\tau^{5}\tau^{6}\tau^{7} = - \,\one_8
 \label{chiral7}
\end{equation}
where $\one_8$ is the $(8\times 8)$ identity matrix.

\subsection{$\mathrm{SO}(8)$ $\Gamma$-matrices}
\label{subsec:gamma}
The eight $\Gamma$-matrices of $\mathrm{SO}(8)$, satisfying $\big\{\Gamma^\mu,\Gamma^\nu\big\}=2\delta^{\mu\nu}$, can be described as
\begin{equation}
  \Gamma^i=\ii\tau^i\otimes \sigma^1\quad,\quad 
  \Gamma^8= -\one_8\otimes \sigma^2
\label{gamma8}
\end{equation}
where the $\sigma$'s are the usual Pauli matrices and the $\tau$'s are the seven matrices defined
in (\ref{gamma7}). This is a Weyl representation, since the chirality matrix is
\begin{equation}
\Gamma\equiv\Gamma^1\Gamma^2\Gamma^3\Gamma^4\Gamma^5\Gamma^6\Gamma^7\Gamma^8 =
\one_8\otimes\sigma^3~.
 \label{gamma9}
\end{equation}
Note that in this realization all $\Gamma$-matrices are anti-symmetric.
{From} (\ref{gamma8}) we easily find that the commutators of two $\Gamma$'s are
\begin{equation}
\label{commgamma1}
\Gamma^{ij}\equiv \frac{1}{2}\big[\Gamma^i,\Gamma^j\big]= -\tau^{ij} \otimes \one_{2}\quad,\quad
\Gamma^{i8}\equiv \frac{1}{2}\big[\Gamma^i,\Gamma^8\big]= \tau^{i}\otimes \sigma^3~;
\end{equation}
thus they all are block diagonal, namely
\begin{equation}
\Gamma^{\mu\nu}= \begin{pmatrix} \gamma^{\mu\nu} & 0 \\ 0
& \bar\gamma^{\mu\nu}\end{pmatrix}
\label{block}
\end{equation}
with the chiral blocks ($\alpha,\beta=1,\ldots,8$) given by
\begin{equation}
(\gamma^{ij})_{\alpha\beta}= -(\tau^{ij})_{\alpha\beta}=\delta^{ij}_{\alpha\beta}+C^{-\,ij}_{\alpha\beta}
\quad,\quad (\gamma^{i8})_{\alpha\beta}=(\tau^i)_{\alpha\beta}=\delta^{i8}_{\alpha\beta}
+C^{-\,i8}_{\alpha\beta}~,
 \label{gamma+}
\end{equation}
and the anti-chiral blocks ($\dot\alpha,\dot\beta=1,\ldots,8$) given by
\begin{equation}
(\bar\gamma^{ij})_{\dot\alpha\dot\beta}= -(\tau^{ij})_{\dot\alpha\dot\beta}
=\delta^{ij}_{\dot\alpha\dot\beta}+C^{-\,ij}_{\dot\alpha\dot\beta}
\quad,\quad (\bar\gamma^{i8})_{\dot\alpha\dot\beta}
=-(\tau^i)_{\dot\alpha\dot\beta}=-\delta^{i8}_{\dot\alpha\dot\beta}
-C^{-\,i8}_{\dot\alpha\dot\beta}~.
 \label{gamma-}
\end{equation}
Notice that the chiral matrices $\gamma^{\mu\nu}$ can be written in a covariant way
with respect to both the vector and the spinor indices, namely
\begin{equation}
(\gamma^{\mu\nu})_{\alpha\beta}= \delta^{\mu\nu}_{\alpha\beta}+C^{-\,\mu\nu}_{\alpha\beta}~.
 \label{gamma+1}
\end{equation}
On the contrary, this is not possible for $\bar\gamma^{\mu\nu}$. 
However, by using the octonionic tensor $C^+$, and splitting the anti-chiral
spinor indices as $\dot\alpha=(\dot a,\dot 8)$, one can show that
\begin{equation}
\label{commgamma3}
(\bar\gamma^{\mu\nu})_{\dot a \dot b} 
= \delta^{\mu\nu}_{\dot a \dot b}-C^{+\mu\nu}_{\dot a \dot b}
\quad,\quad
(\bar\gamma^{\mu\nu})_{\dot a \dot 8}  = -\delta^{\mu\nu}_{\dot a \dot 8}
+C^{+\mu\nu}_{\dot a \dot 8}~.
\end{equation}
This formulation is covariant only in the vector indices but not in the spinor indices.
Eq.s~(\ref{gamma+1}) and (\ref{commgamma3}) are useful to establish a connection between the
matrices $\gamma^{\mu\nu}$ and $\bar\gamma^{\mu\nu}$
and the projection operators $P_1^\pm$ and $P_2^\pm$ defined in
(\ref{P12}). Indeed, one has
\begin{equation}
\begin{aligned}
 \frac{1}{4}(\gamma^{\mu\nu})_{\alpha\beta} & = 
\big(P^-_2-P^-_1\big)^{\mu\nu}_{\alpha\beta} ~,\\
\frac{1}{4}(\bar\gamma^{\mu\nu})_{\dot a\dot b} & = 
\big(P^+_2-P^+_1\big)^{\mu\nu}_{\dot a\dot b} ~,\\
\frac{1}{4}(\bar\gamma^{\mu\nu})_{\dot a\dot 8} & =- 
\big(P^+_2-P^+_1\big)^{\mu\nu}_{\dot a\dot 8} ~.
\end{aligned}
 \label{gammaP12}
\end{equation}
Finally, we recall that the matrices $\gamma^{\mu\nu}$ satisfy the following relations
\begin{equation}
\begin{aligned}
&\gamma^{\mu_1\mu_2}\,\gamma^{\mu_3\mu_4} =
\delta^{\mu_1\mu_4}\delta^{\mu_2\mu_3}-\delta^{\mu_1\mu_3}\delta^{\mu_2\mu_4}
+ \frac{1}{2}\,[\gamma^{\mu_1\mu_2},\gamma^{\mu_3\mu_4}]
+ \gamma^{\mu_1\mu_2\mu_3\mu_4} 
~,\\
&[\gamma^{\mu_1\mu_2},\gamma^{\mu_3\mu_4}] =
2\,\delta^{\mu_2\mu_3}\,\gamma^{\mu_1\mu_4} 
+ 2\,\delta^{\mu_1\mu_4}\,\gamma^{\mu_2\mu_3}
- 2\,\delta^{\mu_2\mu_4}\,\gamma^{\mu_1\mu_3}
-2\,\delta^{\mu_1\mu_3}\,\gamma^{\mu_2\mu_4}~,\\
&\gamma^{\mu_1\mu_2\mu_3\mu_4} = + \frac{1}{4!}\,\epsilon_8^{\mu_1\cdots\cdots\mu_8}\,
\gamma_{\mu_5\mu_6\mu_7\mu_8}~.
\end{aligned}
\label{rel00}
\end{equation}
Similar relations hold for the matrices $\bar\gamma^{\mu\nu}$, but with $\epsilon_8$ replaced by $-\epsilon_8$ in the last one.
Furthermore, we have
\begin{equation}
\label{dualgamma41}
(\gamma^{\mu\nu})_{\alpha\beta}\,(\gamma^{\mu\nu})_{\gamma\delta} 
= 8\,\delta_{\alpha\beta,\gamma\delta}
\quad,\quad
(\bar\gamma^{\mu\nu})_{\dot\alpha\dot\beta}\,(\bar\gamma^{\mu\nu})_{\dot\gamma\dot\delta} 
= 8\,\delta_{\dot\alpha\dot\beta,\dot\gamma\dot\delta}~.
\end{equation}

\section{The $t_8$ tensor}
\label{app:t8}

The explicit expression of the totally anti-symmetric 8-index tensor $t_8$ can be read from
the definition (\ref{f4}). It turns out that $t_8$ can be written as 
\begin{equation}
\label{t8one}
t_8 = \frac{1}{2}\big(T_{(1)} -T_{(2)}\big)
\end{equation}
where $T_{(1)}$ and $T_{(2)}$ are the following single and double trace anti-symmetric tensors
\begin{equation}
 \begin{aligned}
  T_{(1){\phantom{\cdots}}\mu_5\cdots\mu_8}^{\,\mu_1\cdots\mu_4}  &=  
\delta^{[ \mu_2}_{[ \mu_5}\delta^{[ \mu_3}_{\mu_6 ]}\delta^{ \mu_4 ]}_{[ \mu_7}\delta^{ \mu_1]}_{ \mu_8]} ~+~\mbox{permutations}~,
\\
T_{(2){\phantom{\cdots}}\mu_5\cdots\mu_8}^{\,\mu_1\cdots\mu_4}  &=  
\delta^{{\phantom{[}}\!\mu_1 \mu_2}_{{\phantom{[}}\!\mu_5 \mu_6} ~
\delta^{{\phantom{[}}\!\mu_3 \mu_4}_{{\phantom{[}}\!\mu_7 \mu_8}~+~\mbox{permutations}~.
 \end{aligned}
\label{t1t2}
\end{equation}
In $d=8$ the tensor $t_8$ can be given another representation, as explained for example in
Appendix 9.A of Ref. \cite{Green:1987mn}. Let us introduce 
the chiral and anti-chiral zero-mode operators $S_0^\alpha$ and ${\bar S}_0^{\dot\alpha}$
such that
\begin{equation}
\begin{aligned}
\mathrm{Tr}_{S_{0}}\big(S_{0}^{\alpha_1 \alpha_2}S_{0}^{\alpha_3 \alpha_4}S_{0}^{\alpha_5 \alpha_6}
S_{0}^{\alpha_7 \alpha_8}\big) &= \epsilon_8^{\alpha_1\cdots\cdots\alpha_8} ~,\\
\mathrm{Tr}_{{\bar S}_0}
\big({\bar S}_0^{\dot\alpha_1\dot\alpha_2}{\bar S}_0^{\dot\alpha_3\dot\alpha_4}{\bar 
S}_0^{\dot\alpha_5\dot\alpha_6}{\bar S}_0^{\dot\alpha_7\dot\alpha_8}\big) &= \epsilon_8^{\dot\alpha_1\cdots\cdots\dot\alpha_8}~,
\end{aligned}
\label{trS0}
\end{equation}
and define the bi-linear operators
\begin{equation}
 R^{\mu\nu} = \frac{1}{4} (\gamma^{\mu\nu})_{\alpha\beta} S_0^\alpha 
S_0^\beta \quad,\quad
\bar R^{\mu\nu} = \frac{1}{4} (\bar\gamma^{\mu\nu})_{\dot\alpha\dot\beta} {\bar S}_0^{\dot\alpha} 
{\bar S}_0^{\dot\beta} ~.
\label{R+-}
\end{equation}
Then let us consider the following anti-symmetric 8-index tensors
\begin{equation}
\begin{aligned}
t^{\mu_1\mu_2\cdots\mu_7\mu_8}_{+} &= {\phantom{\frac12}}\!\!\!\mathrm{Tr}_{S_0}
\big(R^{\mu_1\mu_2} R^{\mu_3\mu_4}R^{\mu_5\mu_6}R^{\mu_7\mu_8}\big)\\ 
&= \frac{1}{2^8}(\gamma^{\mu_1\mu_2})_{\alpha_1\alpha_2}
(\gamma^{\mu_3\mu_4})_{\alpha_3\alpha_4}
(\gamma^{\mu_5\mu_6})_{\alpha_5\alpha_6}
(\gamma^{\mu_7\mu_8})_{\alpha_7\alpha_8}\,\epsilon_8^{\alpha_1\cdots\cdots\alpha_8}~,\\
t^{\mu_1\mu_2\cdots\mu_7\mu_8}_{-} &= {\phantom{\frac12}}\!\!\!\mathrm{Tr}_{\bar S_0}
\big(\bar R^{\mu_1\mu_2}
\bar R^{\mu_3\mu_4}\bar R^{\mu_5\mu_6}\bar R^{\mu_7\mu_8}\big) \\
&=\frac{1}{2^8}(\bar\gamma^{\mu_1\mu_2})_{\dot\alpha_1\dot\alpha_2}
(\bar \gamma^{\mu_3\mu_4})_{\dot\alpha_3\dot\alpha_4}
(\bar \gamma^{\mu_5\mu_6})_{\dot\alpha_5\dot\alpha_6}
(\bar \gamma^{\mu_7\mu_8})_{\dot\alpha_7\dot\alpha_8}
\,\epsilon_8^{\dot\alpha_1\cdots\cdots\dot\alpha_8}~.
\end{aligned}
\label{t+t-}
\end{equation}
Being completely anti-symmetric invariant tensors of $\mathrm{SO}(8)$, both $t_+$ and $t_-$ 
must be linear combinations of $\epsilon_8$ and of the single and double trace tensors
$T_{(1)}$ and $T_{(2)}$ defined in (\ref{t1t2}). Indeed, using the explicit expressions
(\ref{gamma+1}) and (\ref{commgamma3}) for the matrix elements of $\gamma^{\mu\nu}$
and $\bar\gamma^{\mu\nu}$, with straightforward algebra one can prove that
\begin{equation}
t_\pm = \pm\frac{1}{2}\,\epsilon_8 +\frac{1}{2}\,T_{(1)}-\frac{1}{2}\,T_{(2)}~,
 \label{t+-1}
\end{equation}
from which, using (\ref{t8one}) it follows that
\begin{equation}
t_8 = \mp\frac{1}{2}\,\epsilon_8 + t_\pm ~.
 \label{t88}
\end{equation}
By using the relations (\ref{gammaP12}) in (\ref{t+t-}), we can obtain yet 
another representation of the $t_8$ tensor, namely
\begin{equation}
\begin{aligned}
 t_8^{\mu_1\mu_2\cdots\mu_7\mu_8} &= -\frac{1}{2}\,\epsilon_8^{\mu_1\cdots\cdots\mu_8}
+\big(P^-_2-P^-_1\big)^{\mu_1\mu_2}_{\alpha_1\alpha_2}\cdots
\big(P^-_2-P^-_1\big)^{\mu_7\mu_8}_{\alpha_7\alpha_8}\,\epsilon_8^{\alpha_1\cdots\cdots\alpha_8}~,\\
t_8^{\mu_1\mu_2\cdots\mu_7\mu_8} &= +\frac{1}{2}\,\epsilon_8^{\mu_1\cdots\cdots\mu_8}
-\big(P^+_2-P^+_1\big)^{\mu_1\mu_2}_{\dot\alpha_1\dot\alpha_2}\cdots
\big(P^+_2-P^+_1\big)^{\mu_7\mu_8}_{\dot\alpha_7\dot\alpha_8}\,\epsilon_8^{\dot\alpha_1\cdots\cdots
\dot\alpha_8}~.
\end{aligned}
 \label{t8P12}
\end{equation}
Finally, by exploiting the triality among the $\mathrm{SO}(8)$ representations, we can write also
the following relations
\begin{equation}
\label{t8three}
t_8^{\mu_1\mu_2\cdots\mu_7\mu_8} = \mp \frac{1}{2}\,
\epsilon_8^{\mu_1\cdots\cdots\mu_8}
\pm \big(P^\mp_2-P^\mp_1\big)^{\mu_1\mu_2}_{\nu_1\nu_2}\cdots
\big(P^\mp_2-P^\mp_1\big)^{\mu_7\mu_8}_{\nu_7\nu_8}\,\epsilon_8^{\nu_1\cdots\cdots\nu_8}~.
\end{equation}

There are also useful identities involving the tensors $t_\pm$ and the $\gamma$-matrices. In particular
one has
\begin{equation}
t_+^{\mu_1\cdots\cdots\mu_8}\,\gamma_{\mu_5\mu_6}\gamma_{\mu_7\mu_8} =0
\quad,\quad
t_-^{\mu_1\cdots\cdots\mu_8}\,\bar\gamma_{\mu_5\mu_6}\bar\gamma_{\mu_7\mu_8} =0
\label{tgg}
\end{equation}
which follows from (\ref{t+t-}) and (\ref{dualgamma41}).

\section{$\theta$-function identities}
\label{app:theta}

Let us consider the Riemann identity of the Jacobi $\theta$-functions%
\footnote{For the Jacobi $\theta$-functions we use the standard conventions given, for instance,
in \cite{Green:1987mn}.}, which, in the notation of the main text, is
\begin{equation}
\label{i6}
\sum_{a=2}^4 s_a 
\,\theta_a(z|\ii t)  \prod_{i=1}^{3} \theta_a(z_i|\ii t)=
\theta_1(z|\ii t) \prod_{i=1}^{3} \theta_1(z_i|\ii t) - 2\,
\theta_1(z'|\ii t) \prod_{i=1}^{3} \theta_1(z'_i|\ii t)
\end{equation}
where $s_2=-1$, $s_3=1$, $s_4=-1$, and
\begin{equation}
 \begin{aligned}
  z'_1 & = \frac{1}{2}\,(-z_1+z_2+z_3+z) \quad,\quad z'_2 = \frac{1}{2}\,(-z_2+z_1+z_3+z)~,\\
  z'_3 & = \frac{1}{2}\,(-z_3+z_1+z_2+z) \quad,\quad z' = \frac{1}{2}(-z+z_1+z_2+z_3)~.
 \end{aligned}
\label{z'}
\end{equation}
When computed at $z_1=z_2=z_3=0$, the identity (\ref{i6}) reduces to
\begin{equation}
\label{i7}
\sum_{a=2}^4 s_a\,\theta_a(z|\ii t)\,
\theta_a(0|\ii t)^3 = 2 \,\theta_1({z}/{2}|\ii t)^4~.
\end{equation}
By taking (multiple) derivatives of this identity with respect to $z$, we can obtain
other identities. For instance, applying the second derivative, we get
\begin{equation}
\label{i8}
\sum_{a=2}^4 s_a \,\theta_a^{(2)}(z|\ii t)
\theta_a(0| \ii t)^3
= 6\,\theta_1({z}/{2}|\ii t)^2\,\theta_1'({z}/{2}|\ii t)^2 + 2\,
\theta_1(z/2|\ii t)^3\,\theta_1^{(2)}(z/2|\ii t)~,
\end{equation}
which at $z=0$ reduces to 
\begin{equation}
\label{i8a}
\sum_{a=2}^4 s_a \,\theta_a^{(2)}(0|\ii t)\,\theta_a(0|\ii t)^3
= 0~.
\end{equation}
This is the identity (\ref{idtheta2}) used in the main text.
Analogously, by considering the fourth derivative of (\ref{i6}) we get 
\begin{equation}
\label{i9}
\sum_{a=2}^4 s_a \,\theta_a^{(4)}(z|\ii t)\,\theta_a(0| \ii t)^3
= 3\,\theta'_1(z/2|\ii t)^4 + \theta_1(z/2|\ii t) \,\big[\cdots\big]~,
\end{equation}
which at $z=0$ becomes
\begin{equation}
\label{i9a}
\sum_{a=2}^4 s_a \,\theta_a^{(4)}(0|\ii t)\,\theta_a(0|\ii t)^3
=3\,\theta'_1(0|\ii t)^4
\end{equation}
namely the identity (\ref{idtheta4}) of the main text.

\providecommand{\href}[2]{#2}\begingroup\raggedright

\endgroup

\end{document}